# Strategic forecasting of internet of things technologies through patent social network and innovation cluster analysis

Mehrdad Maghsoudi[1], Reza Nourbakhsh[2], Mehrdad Agha Mohammad Ali Kermani[3*] and Rahim Khanizad[3]

*Correspondence:
Mehrdad Agha Mohammad Ali Kermani
m_kermani@iust.ac.ir
[1]Department of Industrial and Information Management, Faculty of Management and Accounting, Shahid Beheshti University, Tehran, Iran
[2]Iran University of Science and Technology, Tehran, Iran
[3]Faculty of Economy and Management, Iran University of Science and Technology, Tehran, Iran

**Abstract**
The rapid proliferation of Internet of Things (IoT) technologies necessitates robust forecasting mechanisms to guide strategic decision-making amid increasingly complex innovation landscapes. Despite extensive research employing patent analysis for technology forecasting, existing studies lack systematic integration of social network analysis, advanced text mining, and life cycle modeling to comprehensively map IoT technological evolution and collaborative dynamics. This study addresses these gaps by analyzing 154,227 IoT-related patents through a unified methodological framework combining BERT-based text embeddings, k-means clustering with Davies-Bouldin optimization, S-curve life cycle modeling, and Louvain community detection. The analysis identified nine distinct technology clusters spanning foundational infrastructure (Smart Monitoring and Sensor Systems, Network Communication and Data Transmission) to domain-specific applications (Agricultural IoT, Connected Vehicle Technologies). Life cycle assessment revealed temporal convergence, with eight clusters reaching saturation between 2023 and 2027, reflecting ecosystem-wide synchronization driven by standardization imperatives, platform consolidation, and pandemic-accelerated digital transformation. Social network analysis uncovered five major collaborative communities exhibiting divergent strategic orientations: from extreme specialization (Global Telecommunications Technology Leaders: 87.7% network communication focus) to diversified portfolios (China State Grid IoT Consortium: balanced infrastructure investment). Cross-analysis revealed complementary innovation strategies where infrastructure operators pursue breadth, telecommunications specialists maintain focused expertise, and academic researchers emphasize development-aligned agendas. These findings provide actionable intelligence for corporations navigating saturation domains, SMEs adopting focused positioning strategies, and policymakers optimizing intervention timing. The study advances methodological rigor in patent-based forecasting while delivering practical insights for strategic planning, resource allocation, and partnership formation within the evolving IoT ecosystem.

**Article Highlights**

- Analysis of 154,227 patents reveals nine IoT technology clusters with 8 reaching saturation by 2027.

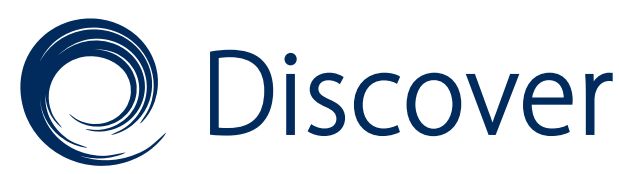

- Network analysis identifies five collaboration communities with distinct specialization patterns across technologies.
- Combined life cycle and network insights enable strategic decisions for corporations, SMEs, and policymakers.

## 1 Introduction

The Internet of Things (IoT) has emerged as a transformative technology, connecting physical objects through seamless data collection, analysis, and transmission to improve quality of life [157]. Since its inception in 1999 [95], IoT has evolved to encompass diverse applications ranging from smart homes and cities to healthcare, agriculture, and industrial automation. With projections indicating over 25 billion connected devices by 2025 [127], IoT offers substantial potential for cost reduction and efficiency improvements across various sectors [93].However, this rapid growth presents significant challenges in network management, load balancing, energy consumption, security, scalability, and fault tolerance [136]—challenges expected to intensify with the advent of 6G and subsequent wireless technologies [120].

In this context, accurately forecasting technological trends has become crucial for businesses to maintain competitive advantage and ensure compliance with industry standards [13]. Technology forecasting aids in managing research endeavors and supports informed decision-making by providing insights into forthcoming technological advancements [70]. Organizations must carefully assess where technologies stand within their life cycles when considering investments and adoption strategies [13, 57].

Despite extensive research on technology forecasting using patent analysis, several critical gaps persist in the IoT domain. First, while numerous studies have employed traditional bibliometric methods to predict technology trends [24, 32], few have systematically integrated social network analysis (SNA), advanced text mining techniques, and patent data analysis to forecast IoT technology trends [8, 99]. Existing studies have predominantly focused on either technological clustering [13, 87] or network analysis [106] in isolation, rather than synthesizing these approaches for a holistic understanding of technological evolution and collaborative dynamics [156].

Second, although recent studies have examined patent networks in related domains such as blockchain [23, 156] and smart cities [87], the IoT field lacks integrated analysis that simultaneously identifies technology clusters, determines their life cycle stages, maps collaboration networks, and examines innovation strategies of key communities [1]. This absence limits our understanding of how collaborative groups contribute to specific technological domains and how these contributions evolve over time [38].

Third, previous research has insufficiently addressed how different organizational types—from large technology corporations to small and medium enterprises (SMEs) and academic institutions—can leverage patent analysis insights for strategic decision-making [52, 68]. Translating analytical findings into actionable managerial implications for diverse stakeholders represents a significant knowledge gap that this study addresses.

To address these gaps, this study pursues the following research questions:

*RQ1* What are the primary technology clusters characterizing the contemporary IoT patent landscape, and how can they be systematically identified and categorized?
*RQ2* At what stage in the technology life cycle does each IoT technology cluster currently reside, and what does this indicate about future development trajectories?
*RQ3* Who are the key players and collaborative communities in IoT patent development, and what network structures characterize their interactions?
*RQ4* How do different collaborative communities allocate their innovation efforts across technology clusters, and what strategic patterns emerge from this analysis?
*RQ5* What strategic implications can be derived from the integrated analysis of technology clusters and collaboration networks for different organizational types operating in the IoT ecosystem?

This study makes several distinctive contributions to the literature on technology forecasting and IoT innovation. First, it introduces a novel methodological integration combining state-of-the-art natural language processing (BERT-based patent embeddings) [89], unsupervised machine learning (k-means clustering with Davies-Bouldin index optimization) [136], technology life cycle modeling (S-curve analysis) [121], and social network analysis (Louvain community detection) [29] within a unified analytical framework. Second, it analyzes a multi-jurisdictional IoT patent corpus sourced from Lens.org, which consolidates records across major patent authorities. Third, the research provides a dual-perspective analysis that simultaneously examines technological evolution and social structures of innovation [136], revealing strategic patterns obscured in single-dimensional analyses—such as how collaborative communities exhibit distinct preferences for specific technology domains and life cycle stages. Fourth, it offers granular, actionable insights tailored to different stakeholders [128], articulating specific strategic implications for large corporations, SMEs, academic researchers, and policymakers based on empirical findings about technology maturity and collaboration patterns. Finally, the identification of emerging technology clusters with substantial growth potential provides timely intelligence for strategic planning and investment decisions in the IoT sector [113].

The remainder of this paper is organized as follows. Section 2 reviews prior work on IoT innovation, patent-based analytics, social network analysis, and text-mining approaches, and positions the study's contributions within this literature. Section 3 explains the research methodology, covering research design; data collection and preparation; text preprocessing and embedding; clustering procedures; technology life-cycle (S-curve) modeling with validation; collaboration-network construction and community detection; and the integrative analysis workflow. Section 4 presents the empirical analyses and supporting figures/tables, including the data overview, clustering analysis, life-cycle modeling and its validation, temporal patterns, collaboration-network characterization, and cross-analysis linking technology domains and collaboration structures. Section 5 offers the discussion, developing managerial and policy implications, acknowledging limitations, and outlining avenues for future research. Section 6 concludes by summarizing the study's contributions and its significance for understanding the evolving IoT innovation landscape.

## 2 Literature review

The accelerating pace of technological innovation in the Internet of Things (IoT) domain has intensified the need for robust foresight mechanisms that can anticipate technological trajectories, identify emergent clusters, and inform strategic decision-making. Patent documents, as formalized manifestations of inventive activity, have increasingly been leveraged as primary data sources for such analyses, offering granular, longitudinal insights into innovation dynamics [43]. However, the analytical frameworks applied to patent data have evolved considerably over the past two decades, shifting from descriptive bibliometrics toward integrative approaches that combine network analysis, machine learning, natural language processing, and life cycle modeling [92]. Despite these advances, significant gaps persist—particularly concerning the simultaneous examination of technological clustering, network-based ecosystem structures, temporal innovation dynamics, and their strategic and policy implications within rapidly evolving domains such as IoT. This literature review critically examines the current state of patent-based forecasting, social network analysis in innovation contexts, technology clustering methodologies, and life cycle modeling, with particular emphasis on IoT technologies. It identifies methodological and empirical shortcomings in prior work and delineates the specific contributions this study makes in addressing those deficiencies.

### 2.1 Patent-Based technological forecasting: methodological evolution and current state

Patent analysis has transitioned from rudimentary citation counting and classification frequency tabulation to sophisticated computational frameworks integrating multiple analytical layers [136]. Early patent-based forecasting relied predominantly on structured metadata—such as International Patent Classification (IPC) codes, citation patterns, and filing trends—to infer technological maturity and competitive positioning [89]. While such approaches provided baseline insights, they were criticized for their inability to capture semantic richness and for their dependence on classification systems that lag behind actual technological evolution [10].

More recent methodologies have sought to overcome these limitations by incorporating unstructured textual data—titles, abstracts, claims, and full descriptions—through natural language processing (NLP) and text mining techniques [105]. Topic modeling approaches, particularly Latent Dirichlet Allocation (LDA) and its derivatives, have been widely applied to identify latent thematic structures within patent corpora [88]. These techniques enable the discovery of emergent technological themes that may not align neatly with formal classification schemes. However, topic models are inherently sensitive to parameter tuning, require substantial preprocessing, and often produce results that lack interpretability without domain expertise [26].

Furthermore, the integration of machine learning classifiers and supervised learning frameworks has enabled the categorization of patents into predefined technological domains or the prediction of future patent activity based on historical patterns [88]. While promising, these approaches frequently suffer from dataset imbalance, limited generalizability across technological contexts, and the risk of overfitting when applied to smaller or rapidly changing fields such as IoT [15]. Critically, few studies have successfully combined textual analysis with network-based and temporal modeling in a unified framework, leading to fragmented insights that fail to capture the multidimensional nature of innovation ecosystems [107].

In the context of IoT, patent-based forecasting efforts have been sparse and methodologically narrow. Existing studies tend to focus either on broad technological trends without domain-specific granularity [25] or on narrow subfields such as blockchain-enabled IoT [136] or logistics applications [33], thereby overlooking the systemic complexity and cross-domain interactions that characterize IoT innovation [41]. Moreover, the majority of these studies rely on relatively small datasets—often fewer than 10,000 patents—which constrains the robustness of statistical inferences and the detection of emergent sub-clusters [15, 25].

### 2.2 Network analysis in innovation ecosystems and technology clustering

Social Network Analysis (SNA) has emerged as a pivotal tool for mapping relational structures within innovation ecosystems, revealing patterns of collaboration, knowledge flows, and technological interdependencies (Mazlumi & Kermani, 2022a). In patent analysis, SNA is commonly applied to construct co-classification networks (where nodes represent IPC codes or keywords, and edges denote co-occurrence), co-inventor networks, co-assignee networks, and citation networks [78]. These network representations facilitate the identification of central technologies, bridging clusters, and isolated technological niches [79].

However, the application of SNA to patent data is not without methodological challenges. First, the choice of network construction method—such as binary versus weighted edges, and the selection of similarity thresholds—can substantially influence the resulting network topology and the communities detected [25]. Second, many studies apply community detection algorithms (e.g., Louvain, Girvan-Newman) without rigorous validation of cluster stability or semantic coherence, leading to clusters that are statistically significant but substantively meaningless (Mazlumi & Kermani, 2022a). Third, temporal dynamics are often neglected; networks are typically analyzed as static snapshots, obscuring the evolution of technological relationships and the emergence or dissolution of clusters over time [136].

Recent work has sought to address some of these limitations. For instance, dynamic network analysis and temporal graph methods have been applied to trace the evolution of collaboration structures and technology diffusion pathways [25, 88]. Nevertheless, these approaches remain underutilized in IoT-specific research. [107] employed co-occurrence IPC networks to map IoT innovation pathways using 32,557 patents, yet their analysis was largely descriptive and did not incorporate strategic or policy-oriented interpretations. Similarly, [25] applied SNA to academic publications rather than patents, limiting the direct applicability of their findings to industrial innovation contexts.

A critical gap in the literature concerns the integration of network-based clustering with technological life cycle stages. While several studies have identified technology clusters within patent datasets, few have systematically assessed whether these clusters correspond to distinct phases of maturity—such as emergence, growth, maturity, or decline—or whether network centrality metrics correlate with innovation leadership or strategic value [78]. This disconnect limits the utility of network analysis for strategic foresight and policy formulation.

### 2.3 Technology life cycle modeling and forecasting in IoT domains

Technology Life Cycle (TLC) models provide a temporal lens through which the evolution of technologies can be understood and forecasted. Grounded in theories of innovation diffusion and industrial dynamics, TLC frameworks posit that technologies progress through identifiable stages—typically characterized as emergence, growth, maturity, and decline—each associated with distinct patterns of patent activity, citation behavior, and competitive dynamics [92].

Patent-based indicators commonly used to infer TLC stages include citation velocity, patent application growth rates, inventor diversity, and technological convergence metrics [78]. For example, rapidly increasing patent filings coupled with high forward citation rates may signal a growth phase, whereas declining application volumes and citation rates may indicate maturation or obsolescence [15]. Despite the theoretical appeal of these indicators, empirical validation remains limited, and threshold values for stage transitions are often arbitrarily defined or context-dependent [88].

Within the IoT domain, TLC modeling has been notably underexplored. While IoT as a whole is generally recognized as being in a growth or early maturity phase [43], there is considerable heterogeneity across sub-domains—such as industrial IoT, smart healthcare, connected vehicles, and smart cities—which likely exhibit different maturity profiles [136]. Few studies have attempted to disaggregate IoT into its constituent technological clusters and assess their respective life cycle stages [41]. This lack of granularity impedes targeted strategic planning and resource allocation.

Moreover, existing TLC studies rarely incorporate network-based insights. For instance, whether technologies occupying central positions in co-classification networks exhibit different life cycle dynamics compared to peripheral technologies remains an open empirical question [107]. Similarly, the interplay between technological clustering and temporal evolution—such as whether certain clusters emerge earlier and mature faster—has not been systematically investigated [25]. These gaps are particularly problematic for policymakers and firms seeking to identify windows of opportunity for intervention or investment.

### 2.4 IoT innovation landscape: Domains, Risks, and strategic implications

The Internet of Things represents a convergent technological paradigm that integrates sensors, actuators, communication networks, data analytics, and cloud computing to enable real-time interactions between physical and digital systems [43]. Architecturally, IoT systems are typically conceptualized as layered structures encompassing perception layers (sensors/actuators), network layers (gateways, edge computing), and application layers (cloud platforms, analytics) [137]. Since the term's introduction by Kevin Ashton in 1999 [11], IoT has expanded exponentially, with global device counts projected to reach 25 billion by 2030—a 300% increase from 2020 levels [65].

This rapid proliferation has been accompanied by diversification into numerous application domains, including smart cities, healthcare, supply chain management, energy systems, and manufacturing [4]. However, the expansion of IoT also introduces significant technological risks and strategic challenges. Cybersecurity vulnerabilities represent a paramount concern, as the vast attack surface created by billions of interconnected devices exposes critical infrastructures and sensitive data to exploitation [41]. Despite this, patent-based innovation analyses have largely overlooked security-related technological clusters, focusing instead on functional capabilities and application areas [25, 33]. This analytical gap is especially critical given the increasing reliance on AI-enabled risk assessment in IoT, as demonstrated by [131], who developed a dependency-based cyber risk model incorporating explainable AI techniques to assess cascading threats across heterogeneous IoT environments.

Furthermore, the integration of artificial intelligence (AI) and machine learning (ML) into IoT systems—often termed AIoT—has emerged as a frontier area, enabling autonomous decision-making, predictive maintenance, and intelligent resource optimization [3]. Yet, the extent to which AI-IoT convergence is reflected in patent activity, how it clusters within broader IoT innovation networks, and its maturity relative to other IoT domains remain underexamined [105]. This is a notable oversight given the strategic importance of AI-enabled IoT for competitive advantage and policy relevance for regulatory frameworks addressing algorithmic accountability and data governance.

From a policy perspective, IoT innovation poses challenges related to standardization, interoperability, spectrum allocation, privacy regulation, and cross-border data flows [41]. Effective policy foresight requires not only an understanding of technological trajectories but also insights into the organizational and geographic distribution of innovation, the concentration of patenting activity among firms and nations, and the accessibility of foundational versus application-layer technologies [10]. Unfortunately, most existing patent studies in the IoT space provide limited guidance on these strategic and policy dimensions, focusing instead on purely technical characterizations [25, 88]. A recent line of research proposes combining hardware-level security architectures, such as CHERI (Capability Hardware Enhanced RISC Instructions), with AI Bills of Materials (AI BOMs) to enhance system transparency and resilience in AI-driven IoT deployments [132].

### 2.5 Comparative analysis of methodological approaches in related studies

A comparative examination of recent IoT and related technology patent studies reveals both methodological progress and persistent limitations. [24] employed SNA to analyze 7,767 research publications on IoT from 2015 to 2018, demonstrating the field's interdisciplinary nature and growing research activity. While their work provided valuable insights into academic discourse, it did not engage with patent data, thereby excluding the industrial innovation dimension and limiting applicability to technology forecasting and commercialization contexts. Moreover, their use of publication data introduced a temporal lag, as patents typically precede publications in reflecting inventive activity [136].

Chen [32] integrated text mining, Technology Opportunity Analysis (TOA), and Technology-Service Evolution Analysis (TSEA) to explore IoT applications in logistics. Their triangulation of journal articles, patents, and market reports (totaling 403 documents) represented a methodologically diverse approach. However, the relatively small sample size and narrow sectoral focus constrained the generalizability of findings. Furthermore, their study did not employ network-based clustering or life cycle modeling, limiting insights into relational structures and temporal dynamics.

Bamakan [13] applied text mining and clustering to approximately 14,000 blockchain patents, concluding that blockchain technology remained in an emergent phase, with U.S. patents concentrated in financial applications. While their use of a large patent corpus was commendable, the clustering methodology lacked transparency regarding validation procedures, and the absence of network analysis precluded examination of inter-cluster relationships and knowledge flows. Additionally, the study did not assess temporal evolution or life cycle stages systematically, relying instead on aggregate growth trends.

Mazlumi and Kermani [107] analyzed co-occurrence IPC networks of 32,557 IoT patents from 1996 to 2020, offering a network-based perspective on IoT innovation pathways. Their work identified central IPC codes and demonstrated the utility of SNA for mapping technological interdependencies. However, the analysis was primarily structural and descriptive; it did not integrate textual content analysis, nor did it assess life cycle stages or provide strategic recommendations. Furthermore, the reliance solely on IPC co-occurrence may have obscured finer-grained thematic distinctions not captured by formal classification systems.

Kim [87] employed a more integrative methodology—combining topic modeling, community detection, machine learning classification, and multivariate time series analysis—to study 32,020 patents related to smart sustainable cities (SSC). Their approach represented a significant methodological advance, particularly in linking thematic content with temporal patterns. Nevertheless, the study's focus on SSC limited its scope, and the findings may not generalize to the broader IoT landscape. Moreover, while temporal analysis was conducted, explicit life cycle stage classification was not performed, and strategic or policy implications were underdeveloped.

Bhatt [ 23, 136] both applied key-route main path analysis to blockchain patents, tracing technology development trajectories and identifying clusters based on disruptive innovation theory. These studies advanced the understanding of knowledge flows and path dependencies but focused exclusively on blockchain—a sub-domain of IoT—and did not address the broader IoT ecosystem or its diversity of applications and risks [3, 41]. Additionally, neither study integrated network centrality metrics with life cycle assessment or explored policy and strategic dimensions comprehensively.

Collectively, these studies reveal a pattern of fragmentation: methodologies tend to emphasize one or two analytical dimensions (e.g., text mining or network analysis or temporal modeling) while neglecting others. Large-scale, integrative analyses that simultaneously address technological clustering, network structures, life cycle stages, and strategic implications remain rare, particularly in the IoT domain [10, 78]. Furthermore, recent dimensions such as cybersecurity risks, AI integration, and policy foresight have received insufficient attention in patent-based IoT research [3, 65].

### 2.6 Research gaps and study positioning

From the foregoing review, several critical research gaps emerge. First, there is a lack of large-scale, comprehensive patent analyses focused specifically on the IoT domain that integrate multiple methodological layers—text mining, network analysis, clustering, and life cycle modeling—into a cohesive analytical framework [88, 107]. Existing studies either employ smaller datasets, focus on narrow sub-domains, or apply only a subset of relevant techniques, resulting in partial and fragmented insights [15, 25, 33].

Second, the relationship between network-based technological structures and innovation timing remains underexplored. Specifically, it is unclear whether technologies occupying central positions in co-classification or co-occurrence networks correspond to more mature life cycle stages, whether peripheral technologies represent emerging niches, and whether cluster cohesion correlates with innovation velocity or strategic value [78, 136]. This gap limits the ability to derive actionable foresight from network models.

Third, thematic analyses of IoT patents have insufficiently addressed critical emerging issues such as cybersecurity, AI-IoT convergence, and privacy-preserving technologies, despite their growing strategic and policy significance [3, 41, 65]. The absence of dedicated attention to these dimensions in clustering and forecasting frameworks represents a notable empirical and conceptual shortcoming.

Fourth, the translation of analytical findings into strategic and policy-relevant recommendations has been weak. Many studies conclude with technical descriptions of clusters or trends but fail to articulate implications for firms, investors, or policymakers [15, 25]. There is a need for frameworks that bridge analytical outputs with strategic positioning, resource allocation, and regulatory foresight [10, 92].

Finally, existing research has paid limited attention to geographic and organizational dimensions of IoT innovation. Understanding which countries, firms, or institutional actors dominate specific technological clusters, and how this concentration affects accessibility, competition, and diffusion, is crucial for both strategic planning and policy formulation [136]. Yet, few studies integrate such analyses with clustering and life cycle assessments [88].

This study addresses these gaps through a methodologically integrative approach applied to a large-scale dataset of IoT-related patents. By combining keyword-based text mining with co-word network analysis and community detection algorithms, the study identifies and characterizes distinct technological clusters within the IoT landscape. Subsequently, it assesses the life cycle stages of these clusters using patent activity metrics, enabling differentiation between emergent, growing, mature, and declining domains. Furthermore, it examines the structural properties of clusters within the network—such as centrality and modularity—and relates these to innovation dynamics and strategic positioning. Critically, the study extends beyond technical description to offer strategic and policy-oriented insights, particularly concerning cybersecurity, AI integration, and sectoral concentration. In doing so, it responds directly to the identified gaps, advancing both the methodological rigor and practical relevance of patent-based IoT forecasting [43, 65, 105]. The Comparative Analysis of Related Works has been shown in Table 1.

The present study departs from prior research by simultaneously addressing technological clustering, network structure, life cycle assessment, and strategic implications within a large-scale IoT patent dataset. Unlike earlier works that emphasize isolated dimensions or narrow sub-domains, this research integrates multiple analytical techniques to provide a holistic, actionable understanding of IoT innovation dynamics. By linking network positions with temporal maturity and thematic content, it offers novel insights into the structural and evolutionary characteristics of IoT technologies. Furthermore, the study's attention to emerging risks—such as cybersecurity vulnerabilities and AI integration—positions it at the intersection of technological, strategic, and policy-relevant inquiry, thereby advancing both academic discourse and practical foresight in the IoT domain.

**Table 1** Comparative analysis of related works

| Study | Focus | Methods | Dataset Size | Scope | Key Limitations |
|---|---|---|---|---|---|
| Bhattacharya et al. [24] | IoT research structure | SNA | 7,767 publications | Academic discourse | No patent data; temporal lag; no TLC modeling |
| [33] | IoT in logistics | TOA, TSEA, text mining | 403 documents | Logistics sector | Small sample; narrow sectoral focus; no network or TLC analysis |
| Bamakan et al. [13, 14] | Blockchain trends | Text mining, clustering | ~ 14,000 patents | Blockchain | Lack of validation; no network analysis; limited temporal modeling |
| Mazlumi & Kermani (2022) | IoT innovation pathways | SNA (IPC co-occurrence) | 32,557 patents | IoT (broad) | Descriptive; no textual content analysis; no TLC assessment; limited strategic insights |
| [88] | Smart sustainable cities | Topic modeling, ML, time series | 32,020 patents | SSC technologies | Narrow scope; no explicit TLC stages; limited policy discussion |
| Tseng et al. (2023) | Blockchain technology paths | Main path analysis, clustering | Patent data | Blockchain | Focused on blockchain only; no broader IoT context; limited integration of network metrics with TLC |
| [23] | Blockchain evolution | Main path analysis | U.S. patents (1974–2021) | Blockchain | Narrow to blockchain; no broader IoT; limited strategic/policy guidance |

## 3 Methodology

### 3.1 Research design

This study employs a mixed-method approach combining computational text analysis, growth modeling, and network science to examine the Internet of Things (IoT) patent landscape. The research design integrates three complementary analytical components applied sequentially to the same patent dataset: (1) unsupervised machine learning to identify technology clusters, (2) S-curve modeling to assess technology maturity stages, and (3) social network analysis to map collaborative relationships among patent assignees [129]. Figure 1 illustrates the overall research workflow, showing how data flows from collection through preprocessing, parallel analysis streams, and final integration.

As shown in Fig. 1, the methodology begins with systematic patent data collection, followed by preprocessing to prepare texts for analysis. The workflow then splits into two parallel analytical streams: one focusing on technological content (clustering and life cycle analysis) and another on collaborative structures (network analysis). These streams converge in an integrative phase that examines relationships between technology domains and collaboration communities.

This design allows the study to address three core research questions: What are the distinct technological domains within IoT innovation? What is the maturity level of each domain? Who are the key actors and how do they collaborate? By triangulating these perspectives, a comprehensive understanding of both the technical and social dimensions of IoT patent activity can be achieved [156].

### 3.2 Data collection and Preparation

#### *3.2.1 Patent database and search strategy*

Patent data were retrieved from Lens.org, a comprehensive open-access patent database aggregating records from over 100 patent offices worldwide, including the United States Patent and Trademark Office (USPTO), European Patent Office (EPO), and World Intellectual Property Organization (WIPO) [123]. Lens.org was selected based on four criteria that ensure methodological rigor and reproducibility. First, the platform provides free accessibility, enabling other researchers to replicate the study without subscription barriers. Second, it offers global coverage across multiple jurisdictions, reducing geographical bias in patent data. Third, standardized metadata fields across different patent offices facilitate data harmonization and analysis. Fourth, bulk download capabilities enable efficient retrieval of large datasets necessary for comprehensive analysis [126]. While commercial databases such as Derwent Innovation offer enhanced features including sophisticated classification systems and improved family linkage, Lens.org provides sufficient data quality for this study's objectives while ensuring methodological transparency and replicability.

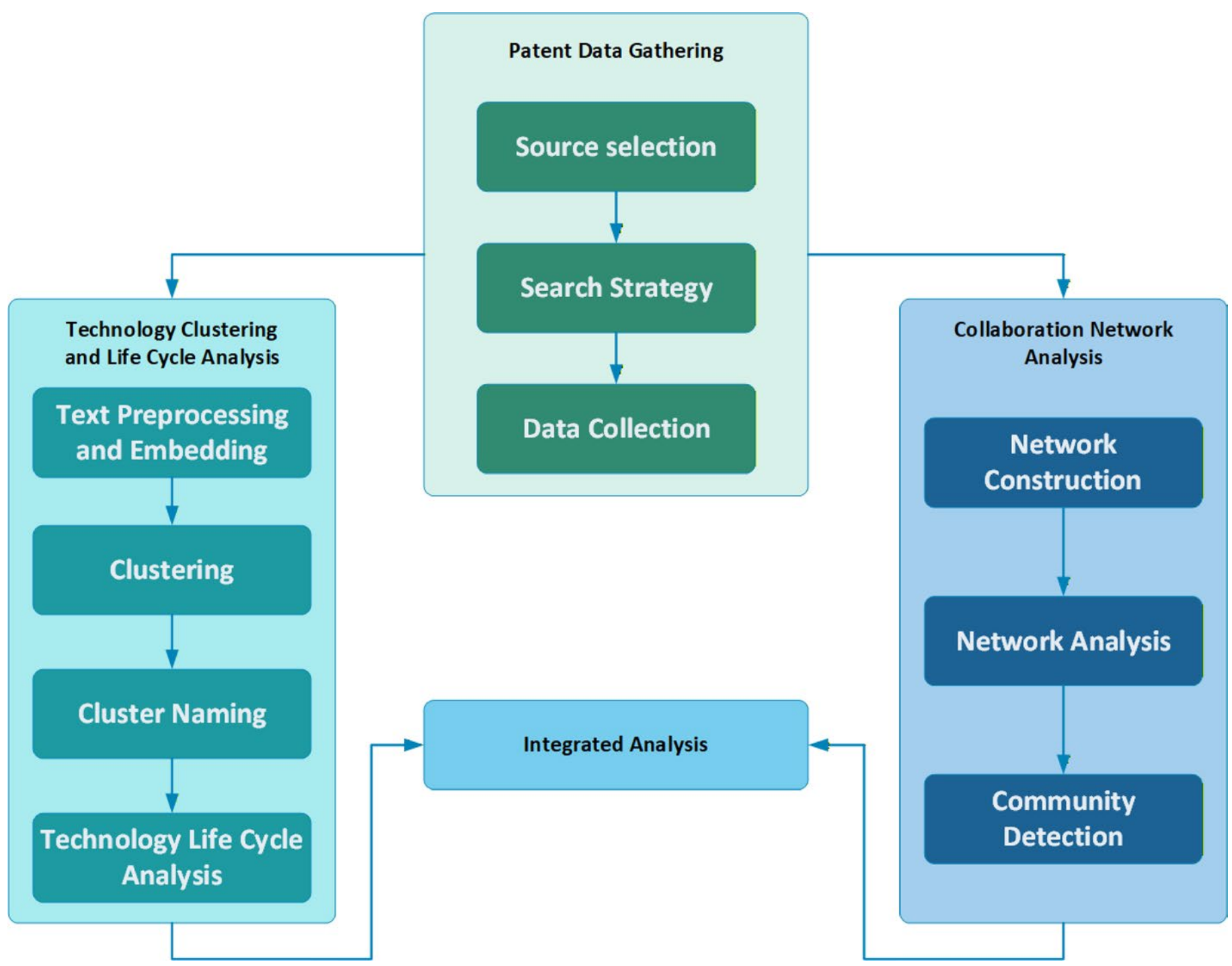

**Fig. 1** Research methodology

The search query applied to the database was:

$$(\text{“Internet of Things” “OR IoT”})\ \text{AND}\ (\text{title OR abstract OR claims})$$

This Boolean search strategy was designed to capture patents explicitly addressing IoT technologies in their core content sections—specifically in titles, abstracts, or claims—rather than merely mentioning these terms in background sections or citations. This focused approach ensures high relevance of retrieved patents to the research domain [136].

The temporal scope of data collection extends from the earliest available IoT patents in the database through October 10, 2025, which represents the date of data extraction for this study. However, a critical methodological decision was made to exclude patents published in 2025 from all quantitative analyses. This exclusion addresses a well-documented bias in patent analysis: publication lag effects. Patent publications typically experience an 18-month delay from filing due to legally mandated secrecy periods, and data from incomplete years create artificial decline patterns that distort trend analyses [60].

Therefore, while patents from 2025 were retrieved and stored, all trend analyses, growth modeling, and temporal statistics utilize data through December 31, 2024. This ensures temporal consistency and avoids right-censoring bias that would otherwise compromise the validity of time-series analyses.

Patent families were not deduplicated in the dataset. This decision reflects the fact that each patent document—regardless of whether it belongs to the same family as another patent—represents a distinct legal instrument with unique claims, assignees, and jurisdictional context [136]. Deduplication could inadvertently remove valuable information about geographical diffusion patterns and multi-jurisdictional protection strategies.

#### *3.2.2 Data fields extracted*

For each patent record in the dataset, the following metadata fields were systematically extracted to support subsequent analyses:

Patent identification information includes the publication number, which serves as a unique identifier for each document in the dataset. Textual content fields comprise the patent title and abstract in full text format, which provide the semantic information necessary for technology classification. Temporal information includes both the filing date (the date when the patent application was submitted) and the publication date (when the patent became publicly available), enabling time-series analysis of innovation trends. Collaborative information encompasses assignee fields, which identify organizations or individuals holding patent rights, and inventor fields, which list individual names of those credited with the invention. Geographic information includes the jurisdictional office (such as USPTO, EPO, or other national offices) where the patent was filed. Technical classification information includes International Patent Classification (IPC) codes where available, providing standardized technology categorization.

Assignee and inventor fields frequently contain multiple entities per patent, recorded as semicolon-delimited strings in the Lens.org export format. These multi-entity fields required systematic parsing and structuring for network construction, as detailed in Sect. 3.5.

#### *3.2.3 Text preprocessing*

Before computational analysis, patent texts underwent systematic preprocessing to remove noise and standardize format. This multi-step process ensures that subsequent analytical methods operate on clean, normalized data rather than raw text with inconsistencies and irrelevant content.

The preprocessing workflow consisted of five sequential steps. First, concatenation was performed by combining the title and abstract fields into a single unified text representation for each patent. This creates a comprehensive textual snapshot of each patent's technological content. Second, normalization was applied by converting all text to lowercase and removing punctuation marks, which eliminates formatting variations that do not carry semantic meaning. Third, tokenization was performed using the Natural Language Toolkit (NLTK) word_tokenize function, which splits continuous text into individual word tokens [27]. Fourth, lemmatization was applied to reduce words to their base forms—for example, converting "connected," "connecting," and "connection" to the common root "connect"—using NLTK's WordNetLemmatizer. This step ensures that morphological variants are treated as equivalent, improving analytical consistency.

Fifth, stop word removal was implemented using a custom stop word list that combines standard English stop words (such as "a," "the," "is," "was") with domain-specific high-frequency terms identified through exploratory analysis (including "internet," "things," "device," "system," "method"). This final step ensures that subsequent embedding and clustering focus on substantive technological content rather than generic terminology [134]. Data quality control required exclusion of records with missing critical fields. Records lacking abstracts were excluded from clustering analysis because abstract content provides essential semantic information for technology classification. Similarly, records with missing assignee or inventor data were excluded from network analysis since collaboration relationships cannot be established without entity information.

### 3.3 Technology clustering analysis

This analytical component aims to identify distinct technological domains within the IoT patent corpus using unsupervised machine learning techniques. The process involves three sequential steps: transforming unstructured text into numerical representations through embedding, grouping similar patents through clustering algorithms, and assigning meaningful interpretive labels to the resulting groups.

#### *3.3.1 Text embedding with BERT*

To capture semantic meaning from patent texts, a transformer-based language model called "bert-for-patents" (anferico/bert-for-patents) was employed. This model represents a domain-adapted variant of BERT (Bidirectional Encoder Representations from Transformers) that has been specifically pre-trained on large corpora of patent documents [89]. BERT models generate contextualized embeddings—numerical vector representations that capture not only individual word meanings but also the relationships between words within their surrounding context [47]. The domain-specific pre-training on patent text enhances the model's sensitivity to technical terminology, legal language patterns, and the unique linguistic conventions found in patent documents [20].

The embedding process transforms each cleaned patent text into a 768-dimensional dense vector—essentially a list of 768 numerical values—that represents its semantic content in a continuous mathematical space. Patents with similar technological content produce vectors that are positioned closer together in this high-dimensional space, while technologically distinct patents yield vectors that are farther apart. This mathematical representation enables computational clustering algorithms to identify groups of semantically related patents.

The selection of BERT-based embeddings over alternative text representation methods—such as Term Frequency-Inverse Document Frequency (TF-IDF), Word2Vec, or Latent Dirichlet Allocation (LDA)—is justified by the superior performance of transformer architectures in capturing complex semantic relationships in technical documents. Traditional bag-of-words approaches like TF-IDF treat words independently and ignore context, while BERT's bidirectional attention mechanism captures nuanced meanings that depend on surrounding words [89]. However, this approach carries limitations: the quality of embeddings depends on the coverage and representativeness of the pre-training corpus, and the high dimensionality (768 dimensions) may introduce computational challenges and potential noise in subsequent clustering steps.

#### *3.3.2 K-Means clustering*

Patent vectors generated through BERT embedding were grouped into clusters using the k-means algorithm, an iterative partitioning method that organizes data points into k distinct groups by minimizing within-cluster variance [103]. Figure 2 illustrates the operational mechanism of the k-means algorithm through its iterative refinement process.

As demonstrated in Fig. 2, the k-means algorithm proceeds through several iterative steps. The process begins with initialization, where k cluster centers (called centroids) are randomly positioned in the data space. In the assignment step, each data point (in this case, each patent vector) is assigned to its nearest centroid based on Euclidean distance, which measures the straight-line distance in high-dimensional space. Following assignment, the update step recalculates each centroid as the geometric mean (average position) of all data points currently assigned to that cluster. These assignment and update steps repeat iteratively until centroids stabilize in position—meaning they no longer move significantly between iterations—or until a predetermined maximum number of iterations is reached [77]. At convergence, the algorithm produces k distinct clusters, each containing patents with similar technological characteristics as captured by their embedding vectors.

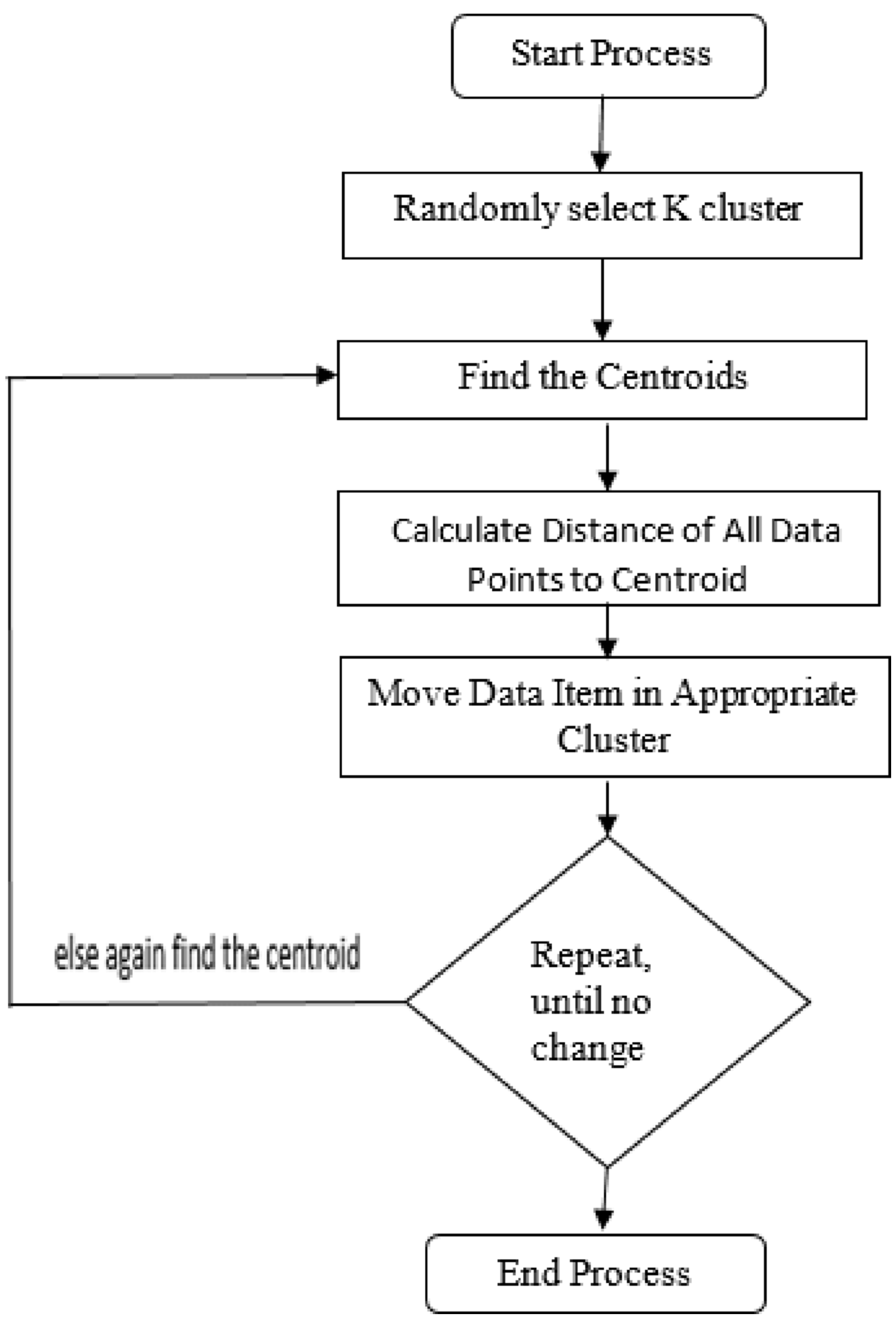

**Fig. 2** The K-means algorithm operation [53]

The selection of k-means over alternative clustering algorithms—including Density-Based Spatial Clustering of Applications with Noise (DBSCAN), Hierarchical DBSCAN (HDBSCAN), Gaussian Mixture Models, and hierarchical clustering—was based on three methodological considerations. First, computational efficiency is critical when analyzing large patent datasets: k-means exhibits near-linear time complexity with respect to the number of data points, making it scalable to datasets containing tens of thousands of patents [136]. Second, interpretability is enhanced by k-means' production of hard cluster assignments, where each patent belongs to exactly one cluster with a clearly defined centroid. This facilitates subsequent labeling and interpretation compared to soft clustering methods that assign probabilistic memberships or density-based methods that may identify noise points not belonging to any cluster. Third, k-means has been extensively validated in patent analysis and technology forecasting literature, providing a methodological foundation for comparison with prior studies [156].

However, acknowledged limitations of k-means include its assumption of spherical cluster shapes (which may not hold in high-dimensional semantic spaces), sensitivity to initialization (though mitigated by k-means++ initialization), and the requirement to prespecify the number of clusters k. Alternative methods such as DBSCAN can automatically determine the number of clusters and handle arbitrary shapes but may struggle with high-dimensional data and require careful parameter tuning. Hierarchical methods provide nested cluster structures but become computationally prohibitive for large datasets.

#### *3.3.3 Determining optimal number of clusters*

A critical methodological decision in k-means clustering is selecting k, the number of clusters to create. Rather than arbitrary selection, a systematic evaluation was conducted across candidate values ranging from $k=2$ to $k=15$ using the Davies-Bouldin Index (DBI) as an internal cluster validation metric [44]. The DBI quantifies cluster quality by measuring the average similarity ratio of each cluster with its most similar cluster, balancing two competing objectives: minimizing within-cluster scatter (making clusters internally compact) and maximizing between-cluster separation (making clusters well-separated from each other). Lower DBI values indicate better-defined cluster structures with tight internal cohesion and clear external boundaries.

The k-means++ initialization algorithm was employed rather than random initialization to improve stability and convergence properties [9]. This method intelligently selects initial centroids by probabilistically favoring points that are far from already-selected centroids, reducing sensitivity to random starting conditions.

The optimal k is determined as the value minimizing the DBI across all tested candidates. However, quantitative optimization metrics alone may not fully capture semantic coherence or practical interpretability of clusters. Therefore, the quantitatively optimal k is cross-validated through qualitative review of cluster contents, examining whether clusters exhibit thematically coherent patent groupings. This combined quantitative-qualitative evaluation ensures both mathematical rigor and semantic meaningfulness in the final cluster structure.

#### 3.3.4 Cluster labeling

After forming clusters through the algorithmic process, descriptive labels are assigned to each cluster to translate mathematical groupings into interpretable technological categories. This labeling process employs a hybrid approach combining quantitative keyword analysis with expert domain knowledge to balance data-driven objectivity with semantic meaningfulness.

The labeling methodology consists of three components. First, keyword extraction is performed for each cluster by identifying the most distinctive terms appearing in member patent abstracts. Term Frequency-Inverse Document Frequency (TF-IDF) scoring is used to weight terms based on their prevalence within a specific cluster relative to the entire corpus [134]. This statistical measure identifies words that are common within a cluster but relatively rare across other clusters, highlighting terminology that characterizes and distinguishes each technology group. Second, representative patent review involves manually examining patents positioned closest to each cluster centroid in embedding space. These central patents typically exemplify the most prototypical technological characteristics of their cluster, providing concrete examples that inform label selection. Third, expert validation engages three domain specialists with complementary backgrounds in IoT technologies, computer science, and engineering. Each expert independently reviews the extracted keywords and representative patents for all clusters, proposes descriptive labels, and participates in consensus-building discussions to resolve labeling discrepancies and ensure semantic validity [67].

This triangulated approach balances the objectivity of computational keyword extraction with the contextual understanding and domain expertise necessary for meaningful interpretation. However, it is acknowledged that cluster labeling inherently involves subjective judgment, and different expert panels might propose alternative labels emphasizing different aspects of cluster content.

### 3.4 Technology life cycle analysis

This component assesses the maturity stage of each identified technology cluster by modeling cumulative patent growth over time using S-curve analysis. The technology life cycle (TLC) framework posits that technologies evolve through predictable developmental stages: emergence, growth, maturity, and saturation [58]. Figure 3 illustrates the theoretical S-curve model that forms the basis for this analysis.

As depicted in Fig. 3, the S-curve divides technology evolution into four distinct stages based on the ratio of current cumulative innovation activity to the carrying capacity (theoretical maximum). The emergence stage, occurring when cumulative patents represent less than 10% of the carrying capacity, is characterized by slow, experimental innovation as foundational concepts are explored and fundamental technical challenges are addressed. The growth stage, spanning approximately 10% to 50% of carrying capacity, exhibits rapid acceleration in innovation activity as the technology gains commercial traction, attracts investment, and demonstrates practical applications. The maturity stage, from 50% to 90% of carrying capacity, shows decelerating growth as fundamental innovations plateau and incremental improvements dominate. The saturation stage, above 90% of carrying capacity, indicates the technology is approaching its natural limits, with minimal new patent activity as the domain becomes fully explored [62].

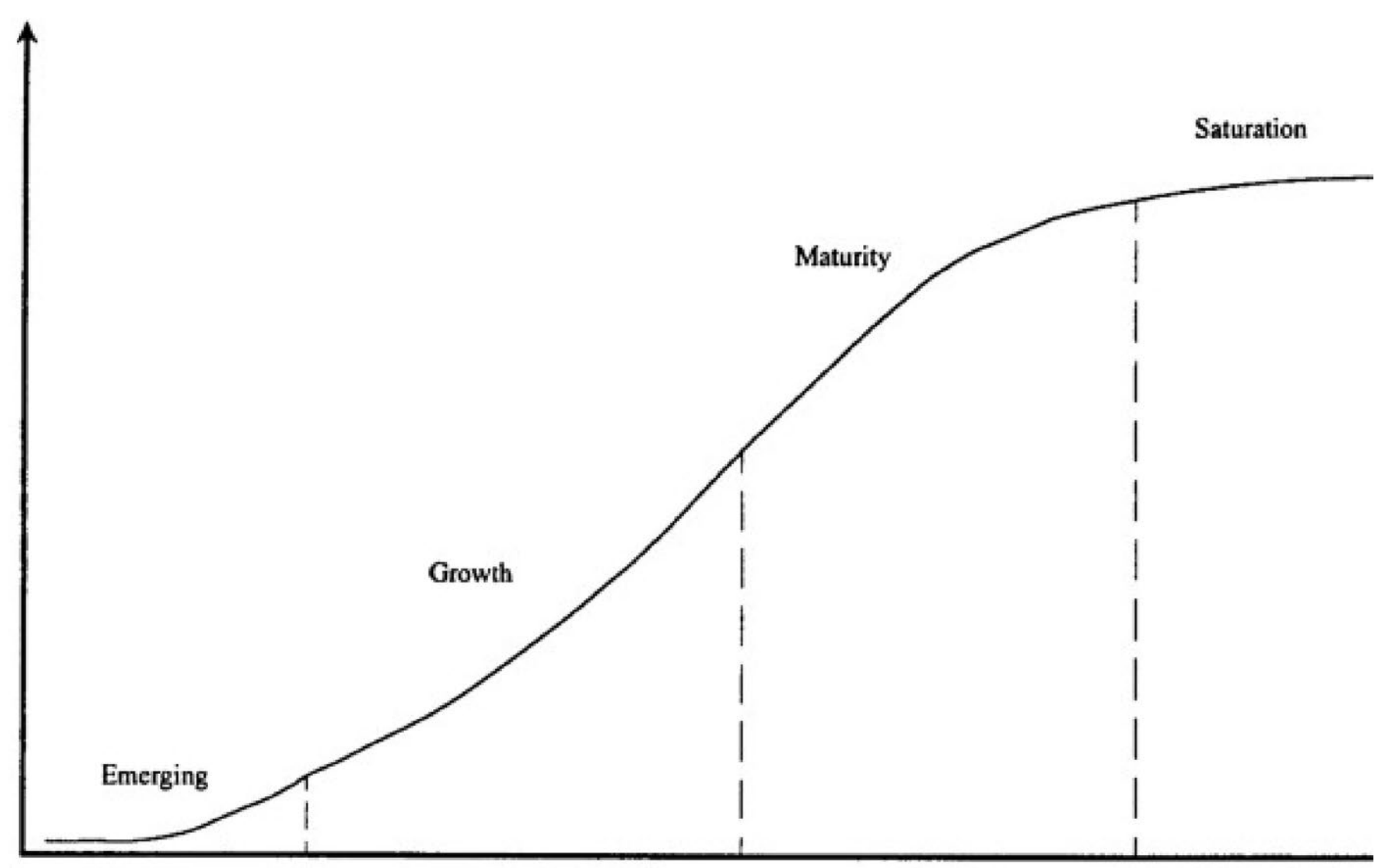

**Fig. 3** The S-curve concept of technology life cycle [62]

#### *3.4.1 Logistic growth model*

Cumulative patent counts over time are modeled using a logistic growth function, which mathematically represents the S-shaped curve pattern:

$$y\left(t\right)=\frac{K}{1+e^{-\frac{(t-a)}{b}}}$$

where $t$ represents time measured in years, $y\left(t\right)$ represents the cumulative number of patents at time $t$, $K$ represents the carrying capacity (asymptotic maximum cumulative patents the technology will reach), $a$ represents the inflection point (the year at which growth rate reaches its maximum value), and $b$ represents the growth rate parameter (controlling the steepness of the curve).

The selection of the logistic growth model over alternative functional forms—such as exponential growth, Gompertz curves, polynomial fits, or Bass diffusion models—is based on both theoretical and empirical considerations. Theoretically, the logistic model aligns with established innovation diffusion theory, which posits that technologies spread through populations following a characteristic S-shaped pattern due to network effects, learning processes, and resource constraints [136]. Empirically, extensive prior research has validated that patent time series in established technological domains frequently exhibit S-shaped growth patterns, making the logistic function an appropriate and well-validated choice for technology forecasting [62]. Interpretively, the model's parameters ( $K, a, b$) have clear substantive meanings—saturation level, timing of maximum growth, and growth speed—that facilitate practical interpretation and comparison across technology domains.

However, several assumptions embedded in the logistic model must be acknowledged. The model assumes unimodal growth with a single inflection point and gradual approach to saturation, which may not hold for technologies experiencing multiple innovation waves or disruptive breakthroughs. The model also assumes stationarity of external conditions, not accounting for exogenous shocks such as policy interventions, breakthrough discoveries in adjacent fields, or market disruptions that could fundamentally alter growth trajectories. For emerging technologies with limited historical data, parameter estimates—particularly the carrying capacity $K$—may exhibit high uncertainty and instability.

#### 3.4.2 Parameter Estimation and curve fitting

For each technology cluster, annual cumulative patent counts are aggregated from the earliest filing date through December 31, 2024. Consistent with the temporal censoring strategy described in Sect. 3.2.1, patents from 2025 are excluded from this time-series analysis to avoid bias from incomplete publication records.

Model parameters ( $K, a, b$) are estimated using nonlinear least-squares regression implemented in SigmaPlot software. This optimization procedure minimizes the sum of squared residuals—the differences between observed cumulative patent counts and model-predicted values:

$$\min_{K,a,b} \sum\, _t [y_{\text{observed}}(t) - y_{\text{model}}(t; K, a, b)]^2$$

The iterative solver requires initial parameter values to begin optimization. These are set based on exploratory data analysis: $K$ is initialized as 1.2 times the observed maximum cumulative count (allowing for potential future growth beyond observed data), a is initialized as the temporal midpoint of the time series, and b is initialized as 0.5 (a small positive value). The solver iteratively adjusts parameters to improve fit until convergence is achieved, defined as parameter changes between iterations falling below a tolerance threshold of $10^{-6}$.

Model fit quality is evaluated using the coefficient of determination ( $R^2$), which measures the proportion of variance in observed data explained by the model. Values of $R^2$ exceeding 0.85 are considered to indicate acceptable fit, following conventions in technology forecasting literature. Visual inspection of residual plots (observed minus predicted values over time) is also performed to identify systematic patterns that might indicate model misspecification.

#### 3.4.3 Life cycle stage classification

Based on estimated model parameters, each technology cluster is classified into one of four life cycle stages using the ratio of 2024 cumulative patents to the estimated carrying capacity $K$ [62]. The classification scheme employs the following thresholds:

Emergence stage:

$$\frac{y\,(2024)}{K} < 0.10$$

indicating that cumulative innovation activity has reached less than 10% of the estimated saturation level. Technologies in this stage are characterized by exploratory research, high uncertainty, and limited commercial deployment.

Growth stage:

$$0.10 \leq \frac{y\,(2024)}{K} < 0.50$$

indicating the technology has achieved between 10% and 50% of saturation. This stage reflects rapid innovation acceleration, increasing commercial interest, and expanding application domains.

Maturity stage:

$$0.50 \leq \frac{y\,(2024)}{K} < 0.90$$

indicating the technology has reached 50% to 90% of saturation. This stage is characterized by slowing growth rates, focus on incremental improvements, and established market positions.

Saturation stage:

$$\frac{y\,(2024)}{K} \geq 0.90$$

indicating the technology has achieved at least 90% of its estimated carrying capacity. This stage suggests limited opportunities for fundamental innovation, with activity focused on optimization and maintenance.

These threshold values follow established conventions in technology forecasting literature but are acknowledged to be somewhat arbitrary. Alternative threshold schemes—such as those based on growth rate derivatives or second derivatives of the S-curve—could yield different stage boundaries.

This analysis is descriptive and exploratory in nature. The S-curve model does not support causal inference regarding the mechanisms driving technology evolution, nor does it enable hypothesis testing about factors influencing growth trajectories. Rather, the model provides a heuristic framework for characterizing current maturity levels and anticipating potential future trajectories under the assumption that historical growth patterns will continue.

### 3.5 Collaboration network analysis

This analytical component examines the social and organizational structures underlying IoT innovation by constructing and analyzing a patent collaboration network. This approach is grounded in social network theory, which conceptualizes innovation as fundamentally relational—shaped by patterns of collaboration, knowledge exchange, and resource sharing among actors embedded in social structures [130].

#### *3.5.1 Network construction*

A collaboration network is constructed as an undirected, weighted graph G = (V, E), where V represents the set of nodes (vertices) and E represents the set of edges (links connecting nodes). Nodes in this network represent unique patent assignees—organizations or individuals holding legal rights to patents. Assignees were selected as the unit of analysis rather than inventors because assignees represent the strategic locus of patent ownership, investment decisions, and commercialization activities [12]. While inventors generate technical knowledge, assignees control exploitation rights and make strategic collaboration decisions, making them more relevant for understanding innovation ecosystems from an organizational and economic perspective.

Edges in the network connect pairs of assignees who appear together as co-assignees on at least one patent, indicating a collaborative relationship. Edge weights quantify the strength of collaboration, defined as the number of distinct patents co-assigned to each pair of nodes. Higher edge weights indicate more intensive or sustained collaborative relationships.

For patents with multiple assignees, all pairwise combinations of assignees are generated to create edges. For example, a patent with three assignees A, B, and C generates three edges: A-B, A-C, and B-C. This approach follows standard practice in co-authorship and co-patenting network analysis [117]. However, it assumes that all co-assignees on a patent engaged in substantive collaboration. In reality, some co-assignments may reflect licensing agreements, corporate acquisitions, or administrative arrangements rather than genuine collaborative R&D partnerships. Distinguishing these different types of relationships is not feasible from patent metadata alone, representing an acknowledged limitation.

Patents with single assignees contain no collaborative information and are therefore excluded from network construction. This exclusion introduces a potential selection bias toward collaborative innovation activities, systematically omitting isolated or proprietary innovation efforts.

#### *3.5.2 Network-Level metrics*

To characterize the global structural properties of the collaboration network, several network-level metrics are computed using Gephi software, an open-source platform for network visualization and analysis [17]. These metrics provide a structural profile of the IoT collaboration landscape, revealing whether innovation activity is concentrated among tightly connected actors or distributed across loosely connected communities.

Network size is the most basic descriptor, consisting of the number of nodes ($|V|$) and number of edges ($|E|$). These values indicate the scale and scope of collaborative activity captured in the dataset.

Network density quantifies how interconnected the network is as a whole, calculated as the ratio of actual edges to the maximum possible edges: $D = \frac{2|E|}{|V|(|V|-1)}$. Density ranges from 0 (no connections exist) to 1 (every possible pair is connected). Low density indicates a sparse, fragmented collaboration landscape with many disconnected or loosely connected actors. High density suggests a tightly interwoven ecosystem where most actors are directly or indirectly connected [136].

Average degree measures the typical breadth of collaboration, calculated as the mean number of direct collaborators per node: $\langle k \rangle = \frac{2|E|}{|V|}$. This metric indicates whether actors typically collaborate with many partners (high average degree) or maintain focused partnerships with few collaborators (low average degree).

Weighted average degree extends this concept by accounting for collaboration intensity rather than just the number of partners. It is calculated as the mean total edge weight per node, revealing whether actors engage in many shallow collaborations or fewer deep collaborations.

Clustering coefficient measures the tendency for network nodes to form tightly-knit groups, quantified as: $C = 3 \times \frac{\text{number of triangles}}{\text{number of connected triples}}$. A triangle exists when three nodes are all mutually connected (A-B, B-C, A-C). High clustering coefficients indicate the presence of cohesive subgroups where collaborators of a given actor also tend to collaborate with each other, forming closed triadic structures [156]. Low clustering suggests more open, bridging collaboration patterns.

calculated as the mean shortest path distance between all pairs of connected nodes: $L = \frac{1}{|V|(|V|-1)} \sum d(i,j)$, where $d(i,j)$ is the shortest path length between nodes $i$ and $j$. Shorter average path lengths suggest rapid diffusion potential, as information can traverse the network through few intermediary steps. Longer path lengths indicate fragmentation or hierarchical structures that may impede knowledge flow [118].

These metrics collectively enable assessment of whether the IoT collaboration network exhibits characteristics of theoretical network archetypes—such as small-world networks (combining high clustering with short path lengths) or scale-free networks (exhibiting power-law degree distributions with hub actors)—which have important implications for innovation dynamics and knowledge diffusion.

#### *3.5.3 Node-Level centrality metrics*

Beyond network-level structure, node-level centrality metrics are computed to identify influential actors occupying strategically important positions within the collaboration network [59]. Different centrality measures capture different dimensions of influence and strategic positioning.

Degree centrality measures the number of direct collaborators for each node, normalized by the maximum possible degree: $C_D(i) = \frac{k_i}{|V|-1}$, where k $k_i$ is the number of edges connected to node $i$. High degree centrality indicates actors with extensive collaborative reach who partner with many different organizations or individuals [59]. These actors may serve as connectors facilitating knowledge exchange across diverse partners.

Weighted degree, also called strength, extends degree centrality by incorporating collaboration intensity: $s_i = \sum w_{ij}$, where the sum is taken over all edges incident to node $i$ and $w_{ij}$ represents edge weight. This metric distinguishes between actors who collaborate broadly with many partners (high degree) versus actors who maintain intensive long-term relationships with select partners (high strength relative to degree).

Betweenness centrality measures the extent to which a node lies on shortest paths between other nodes: $C_B(i) = \sum_{s \neq i \neq t} \frac{\sigma_{st}(i)}{\sigma_{st}}$, where the sum is over all pairs of nodes $s$ and $t$ (excluding $i$), $\sigma_{st}$ is the total number of shortest paths from $s$ to $t$, and $\sigma_{st}(i)$ is the number of those paths passing through node i. High betweenness centrality identifies "brokers" or "bridges" who connect otherwise disparate parts of the network [59]. These actors occupy structural positions that enable them to control or facilitate information flow between different communities or regions of the network, potentially serving as gatekeepers or knowledge brokers.

Closeness centrality quantifies how efficiently a node can access all other nodes in the network, calculated as the inverse of average shortest path distance: $C_C(i) = \frac{|V|-1}{\sum_{j \neq i} d(i,j)}$, where the sum is over all nodes $j \neq i$. High closeness centrality indicates actors positioned to efficiently access or disseminate information across the entire network, requiring few intermediary steps to reach any other actor [19].

These complementary centrality measures reveal different types of influential actors within the collaboration network. Degree and strength centrality identify prolific collaborators with many or intensive partnerships. Betweenness centrality identifies structural brokers connecting otherwise separate groups. Closeness centrality identifies actors with efficient network positions for information access and dissemination. Together, these metrics provide a multidimensional portrait of influence and strategic positioning within the IoT innovation ecosystem.

#### *3.5.4 Community detection*

To uncover meso-scale organizational structures within the collaboration network—clusters of densely interconnected actors that form coherent communities—the Louvain community detection algorithm is applied [29]. This computational method partitions networks into modules or communities characterized by high density of internal edges relative to external edges, revealing the modular organization of collaborative relationships.

The Louvain algorithm operates by optimizing modularity, a quality metric quantifying the density of intra-community edges compared to a null model assuming random edge placement: $Q = \frac{1}{2m} \sum_{i,j} \left[ A_{ij} - \frac{k_i k_j}{2m} \right] \delta \left( c_i, c_j \right)$ ,where $m$ is the total number of edges, $A_{ij}$ is the adjacency matrix indicating presence or absence of an edge between nodes $i$ and $j$, $k_i$ is the degree of node $i$, $c_i$ is the community assignment of node $i$, and $\delta \left( c_i, c_j \right) = 1$ if nodes $i$ and $j$ belong to the same community and 0 otherwise. Higher modularity values indicate stronger community structure.

Figure 4 illustrates the iterative optimization process employed by the Louvain algorithm.

As shown in Fig. 4, the Louvain method proceeds through two phases that repeat iteratively until no further modularity improvement is possible:

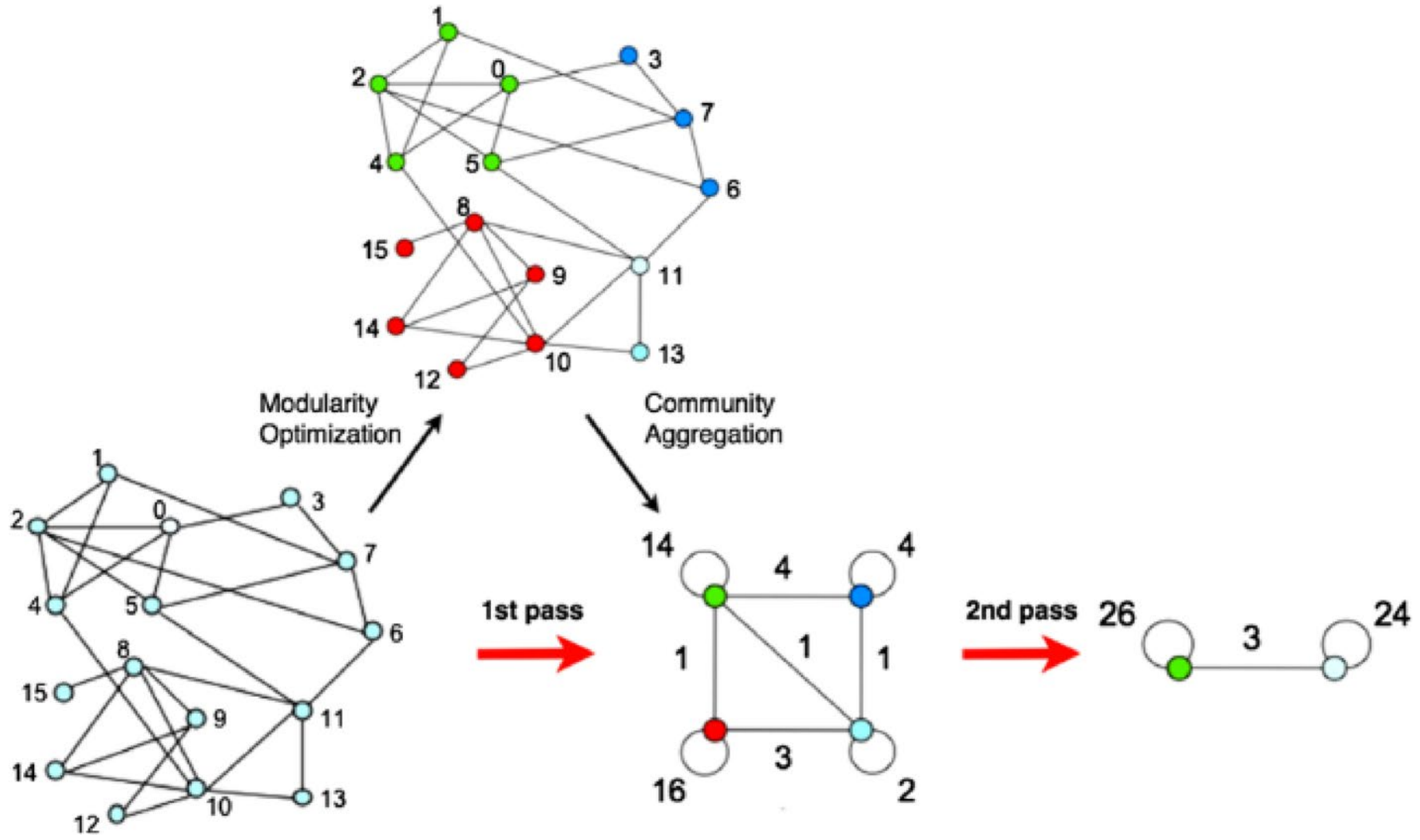

**Fig. 4** Community detection based on increasing modularity [156]

1. *Local Optimization Phase* Each node is initially assigned to its own singleton community. Nodes are then sequentially evaluated to determine whether moving them to a neighboring community (a community containing at least one of their collaborators) would increase overall modularity. If such a move increases modularity, it is accepted; otherwise, the node remains in its current community. This process continues until no single-node moves can further improve modularity.
2. *Aggregation Phase* All nodes belonging to the same community identified in phase one are collapsed into a single "super-node", creating a new network where nodes represent communities rather than individual actors. Edge weights between super-nodes represent the sum of edges between their constituent nodes. These two phases repeat on the newly aggregated network, potentially revealing hierarchical community structures at multiple scales [29]. The algorithm terminates when aggregation no longer improves modularity, yielding a final partition of the original network into communities.

The Louvain algorithm was selected over alternative community detection methods based on several methodological considerations:

- Compared to the Girvan–Newman algorithm, which iteratively removes edges with the highest betweenness centrality, Louvain offers superior computational efficiency for large networks—exhibiting near-linear time complexity rather than polynomial complexity [66].
- Compared to label propagation algorithms, which are faster but less stable and may produce inconsistent results across runs, Louvain provides more deterministic and reproducible community structures [132].
- Compared to Infomap, which optimizes information-theoretic criteria rather than modularity, Louvain aligns more directly with the theoretical goal of identifying densely interconnected modules, though Infomap may be preferable for networks where flow dynamics are of primary interest [136].

Overall, Louvain provides a well-validated balance of computational efficiency, methodological transparency, and alignment with modularity-based theoretical frameworks for community structure.

Following community detection, each identified community is assigned a descriptive label to facilitate interpretation. Labeling is based on examination of the most central actors within each community (those with highest degree or betweenness centrality) or, where applicable, the dominant organizational type or technological focus inferred from assignee names and patent content. This labeling process involves researcher interpretation and is subject to the same subjective limitations discussed for cluster labeling in Sect. 3.3.4.

### 3.6 Integration and synthesis

The final phase of the methodology integrates findings from the three analytical components—technology clustering (Sect. 3.3), life cycle analysis (Sect. 3.4), and network analysis (Sect. 3.5)—to provide a holistic, multidimensional view of the IoT innovation landscape.

For each patent in the dataset, three categorical attributes are assigned through the preceding analyses:

First, technology cluster membership is assigned based on the k-means clustering process, indicating which of the identified technology domains the patent belongs to. Second, the life cycle stage of the technology cluster is assigned based on S-curve modeling, classifying the patent's technology domain as emergence, growth, maturity, or saturation. Third, the community membership of the patent's primary assignee is assigned based on Louvain community detection, indicating which collaborative community the patent's owner belongs to.

These three attributes are cross-tabulated to examine associational patterns and relationships, addressing questions such as:

Which collaboration communities are most active in which specific technology domains? This reveals whether certain organizational networks specialize in particular technological areas or maintain diversified portfolios across multiple domains. Are certain communities concentrated in emerging technologies while others focus on mature technologies? This indicates whether different collaborative networks occupy different strategic positions along the innovation life cycle, potentially reflecting different risk profiles, resource capabilities, or strategic orientations. Do communities exhibit specialized or diversified technology portfolios? This reveals the breadth versus depth of technological focus across different organizational networks.

To visualize these complex many-to-many relationships, Sankey diagrams are constructed showing flows from collaboration communities (source nodes on the left) to technology clusters (target nodes on the right). In these diagrams, the width of each flow is proportional to the number of patents flowing from a particular community to a particular technology cluster [136]. Sankey diagrams are particularly effective for displaying categorical relationships with multiple sources and targets, enabling visual identification of dominant pathways and patterns that would be difficult to discern from tabular data alone.

It is important to emphasize that this integrative analysis is exploratory and descriptive in nature. The observed associations between community membership and technological focus do not establish causal relationships—the analysis cannot determine whether community structure influences technological specialization, whether technological domains shape community formation, or whether both are driven by external factors. Rather, the integration reveals empirical patterns of association that may inform future hypothesis-driven research. The conceptual validity of integrating these analytical dimensions rests on the assumption—supported by prior research in innovation studies—that collaboration structures and technological foci represent interrelated dimensions of innovation ecosystems [130].

## 4 Results

### 4.1 Data gathering

This study employed a systematic approach to collect patent data related to Internet of Things (IoT) technologies using the Lens.org database. Lens.org represents one of the most comprehensive and authoritative global repositories for patent information, providing open access to an extensive archive of patent documentation. The selection of this platform was predicated on its extensive coverage, intuitive interface, and sophisticated search functionalities, which are essential for conducting rigorous research in rapidly advancing technological domains such as IoT.

To ensure comprehensive and pertinent data collection, a structured search strategy was implemented. Two primary keywords were employed: “Internet of Things” and “IoT”, which were strategically selected to capture patents explicitly associated with IoT technologies and their applications. The designated keywords were queried across three critical patent sections: title, abstract, and claims. This methodological approach ensures the identification of patents substantively related to IoT technologies, while filtering out documents with only peripheral references to these concepts. No temporal constraints were imposed on the search query, thereby enabling the capture of the complete historical trajectory of IoT technology development from its inception through October 10, 2025, which serves as the data collection cutoff date for this investigation.

The executed search query on the Lens.org platform yielded a comprehensive dataset comprising 154,227 distinct active patents related to IoT technologies. For each patent record, essential metadata fields were extracted, including patent title, abstract, publication date, inventor information, assignee details, and International Patent Classification (IPC) codes. The title and abstract provide concise technical descriptions of the inventions, which are fundamental for subsequent text mining and clustering analyses. Publication dates facilitate the examination of temporal evolution patterns in IoT technologies, while inventor and assignee data are instrumental for analyzing collaborative networks and identifying principal actors within the field.

To examine the growth trajectory and developmental patterns of IoT patents over time, the distribution of patents was analyzed based on their publication years, as illustrated in Fig. 5. The temporal distribution reveals a pronounced upward trajectory in the annual publication of IoT-related patents, demonstrating the escalating significance and rapid advancement of IoT technologies. A notable acceleration in patent publications is evident commencing around 2015, suggesting a critical inflection point in IoT technology development and commercial adoption. The progression from 79 patents in 2010 to peak activity exceeding 24,000 patents in 2021 underscores the exponential growth in innovation activities. The period between 2015 and 2021 exhibits particularly robust growth, with publication volumes increasing from 2,907 to 24,967 patents. The subsequent years (2022–2024) demonstrate sustained high-volume patent activity, ranging between 17,000 and 22,000 patents annually, indicating a maturation phase while maintaining intensive innovation focus across industries. The data for 2025, representing a partial year through October, shows 10,721 patents, suggesting continued robust activity when annualized. This temporal analysis not only establishes a foundation for understanding the evolutionary trajectory of IoT technologies but also indicates persistent momentum in IoT patent publications, reflecting ongoing innovation and substantial investment in this technological domain. The trends observed in Fig. 5 provide context for subsequent detailed analyses of technology clusters and collaboration networks presented in later sections of this study.

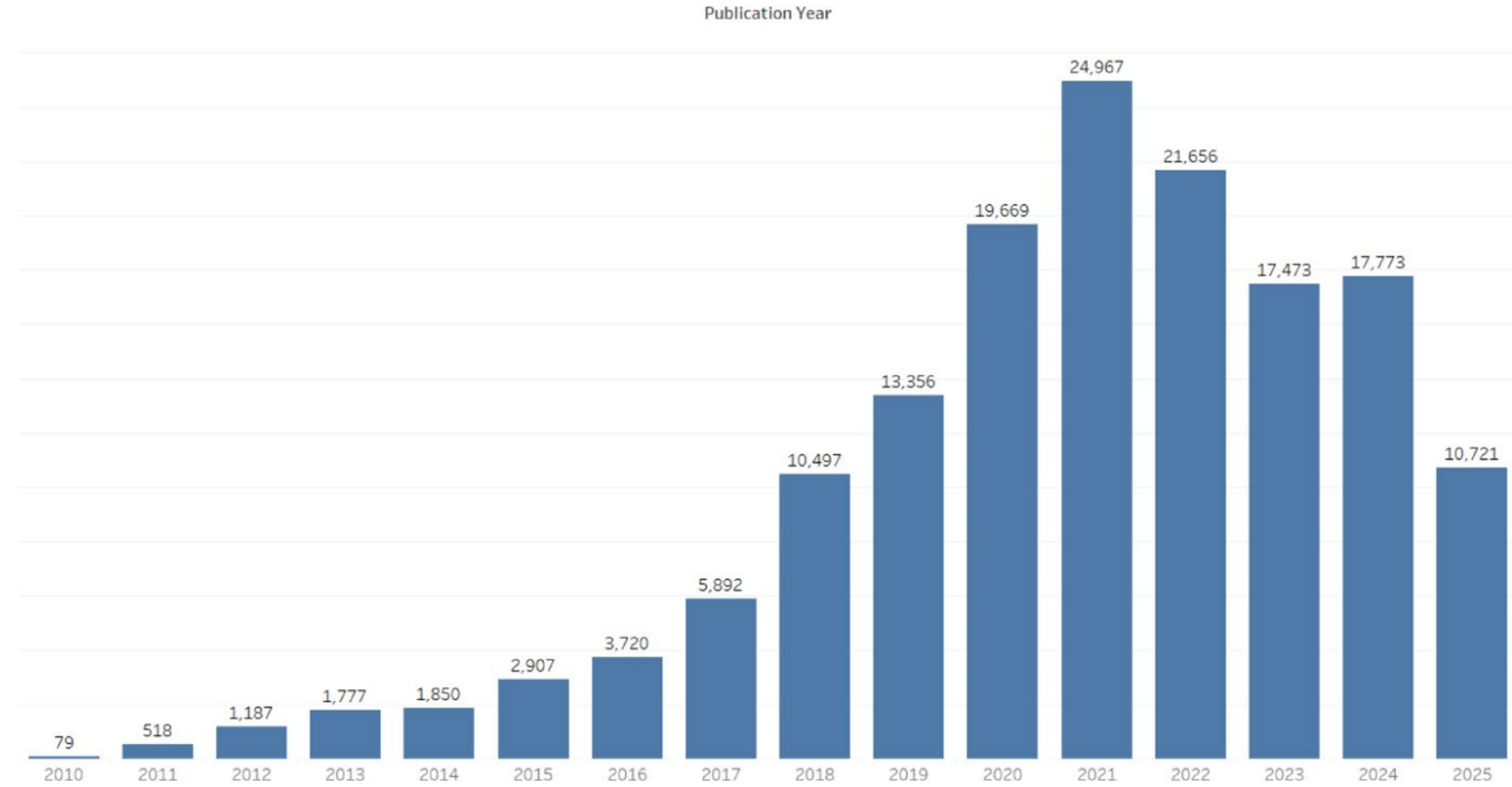

**Fig. 5** Number of patents registered by year

### 4.2 Technology clustering

The technology clustering process in this study encompasses three sequential steps: preprocessing of patent titles and abstracts, patent clustering analysis, and technology life cycle assessment. This systematic approach enables the identification and categorization of distinct technological domains within the IoT patent landscape.

Word embedding represents a fundamental technique in natural language processing (NLP) that transforms textual data into numerical vector representations within a multidimensional vector space, thereby facilitating quantitative text analysis. These numerical vectors are structured as real-valued arrays that encode the semantic meaning of individual words, positioning semantically similar words in proximate locations within the vector space [40, 109]. In this investigation, the titles and abstracts of each patent were converted into n-dimensional vectors of real numbers through word embedding techniques. This transformation process integrates language modeling operations with feature learning methodologies, effectively mapping lexical elements into real number vector representations [28].

The specific text embedding model employed in this study is the "Bert-for-patents" model, which is derived from the BERT (Bidirectional Encoder Representations from Transformers) language model architecture [48]. This model was selected based on its extensive training corpus, encompassing over 100 million patent documents, which provides superior performance in patent text analysis [37]. Figure 6 illustrates the text embedding workflow implemented in this research [102]. The output of this preprocessing phase consists of vector sets representing patent titles and abstracts, which subsequently serve as input for the patent clustering procedure.

For the clustering analysis, the k-means algorithm was employed due to its computational efficiency, robustness, and widespread adoption in patent analysis research [136]. The Davies-Bouldin (DB) index was utilized as the primary evaluation metric to determine the optimal number of clusters. This index quantifies both cluster separation and compactness simultaneously, with lower values indicating superior clustering quality characterized by well-separated and cohesive clusters [44].

To determine the optimal cluster configuration for the patent dataset, the DB index was systematically calculated for k values ranging from 2 to 15 clusters. Table 1 presents the DB index values for each examined cluster number. The analysis reveals a clear pattern wherein the DB index progressively decreases from $k=2$ (DB = 1.847) to $k=9$ (DB = 1.245), after which it begins to increase consistently. The minimum DB index value of 1.245 at $k=9$ indicates the optimal clustering configuration for this dataset. This optimal value of nine clusters suggests a balanced representation of distinct technological domains within the IoT patent landscape, providing sufficient granularity to distinguish between different technology areas while avoiding excessive fragmentation that could result from over-clustering Table 2.

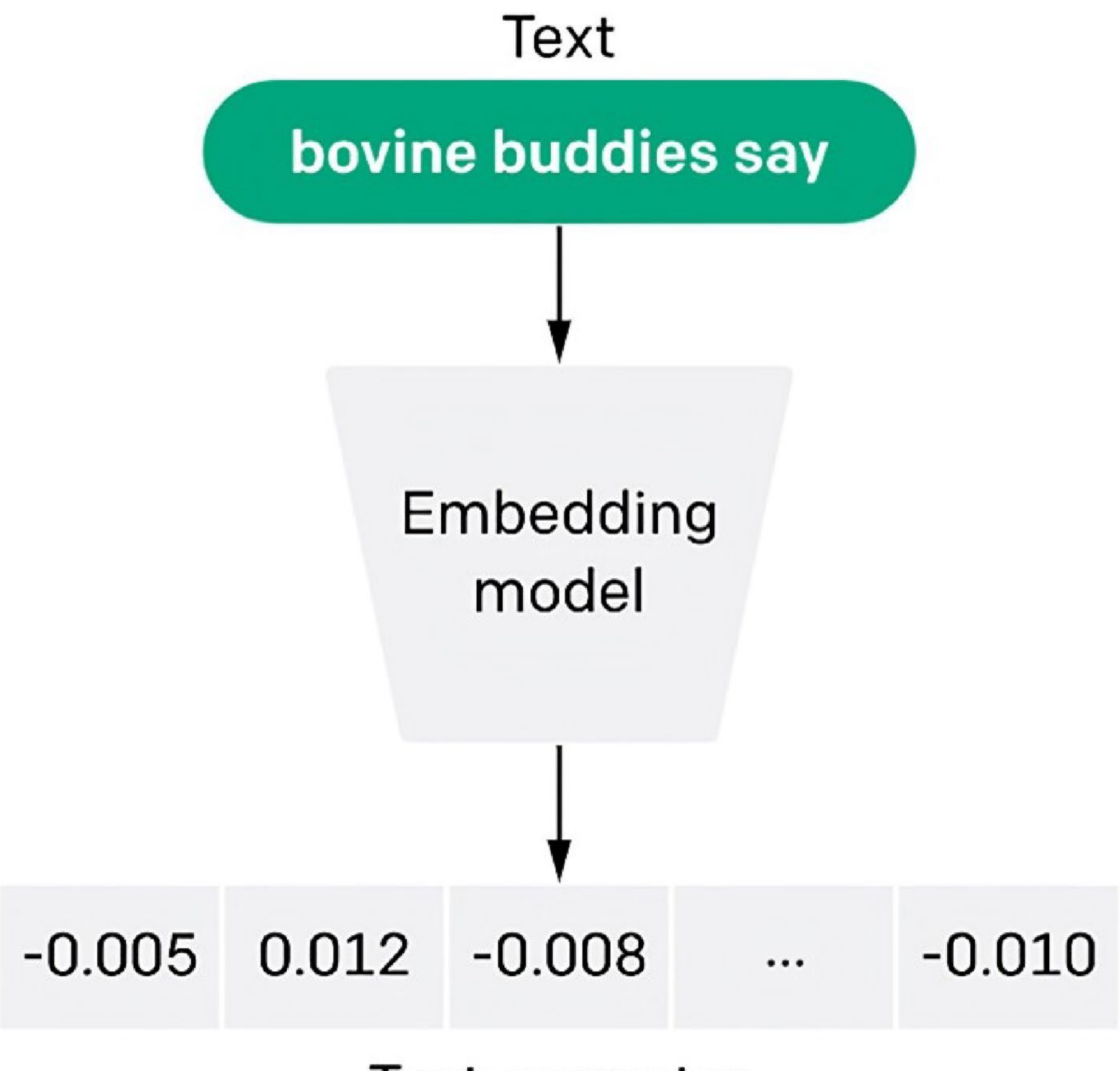

**Fig. 6** Process of text embeddings

**Table 2** Davies-Bouldin index values for cluster optimization

| Number of Clusters (k) | Davies-Bouldin Index | Number of Clusters (k) | Davies-Bouldin Index |
|---|---|---|---|
| 2 | 1.847 | **9** | **1.245** |
| 3 | 1.692 | 10 | 1.263 |
| 4 | 1.578 | 11 | 1.289 |
| 5 | 1.489 | 12 | 1.318 |
| 6 | 1.425 | 13 | 1.352 |
| 7 | 1.368 | 14 | 1.391 |
| 8 | 1.297 | 15 | 1.428 |

### 4.3 Clusters naming

After performing text mining on patent abstracts and titles using unsupervised clustering methodology, nine distinct technological domains within Internet of Things innovation were identified and systematically analyzed. The optimal number of clusters was determined using the Davies-Bouldin index evaluation. Subsequently, term frequency-inverse document frequency (TF-IDF) analysis was employed to extract the most salient keywords from each cluster [136]. The representative keywords characterizing each cluster were selected based on their frequency distribution and contextual relevance to IoT applications.

To assign semantically meaningful and technically accurate labels to each identified cluster, this study employed a modified Delphi method involving a panel of three domain experts specializing in Internet technologies and IoT systems architecture [75, 156]. The Delphi method is a structured communication technique that relies on iterative rounds of expert consultation to achieve convergence toward consensus on complex issues where empirical evidence may be limited [136]. This methodology is particularly appropriate for nomenclatural tasks requiring domain-specific knowledge and interpretive judgment [86].

The naming process proceeded through multiple Delphi rounds. In the initial round, the expert panel was presented with the extracted representative keywords for each cluster and independently proposed cluster labels based on their technical interpretation. In subsequent rounds, experts reviewed the collective responses and refined their suggestions, incorporating feedback from previous iterations [76]. This iterative process continued until convergence was achieved, defined as substantial agreement among all three experts on the cluster nomenclature [49]. The final cluster names were cross-referenced against established IoT literature to ensure technological accuracy and nomenclatural appropriateness. The comprehensive clustering results are presented in Table 3, which establishes the taxonomy of IoT innovation landscapes as derived from patent documentation.

Cluster 0, designated as Smart Monitoring and Sensor Systems, consolidates innovations centered on real-time environmental and operational data acquisition through distributed sensor architectures. The predominance of keywords “module,” “monitor,” and “sensor” reflects foundational IoT infrastructure integrating ubiquitous sensing capabilities with autonomous detection mechanisms. Patents in this cluster address sensor miniaturization, edge-based data collection, and intelligent anomaly recognition systems critical for industrial monitoring applications. The prevalence of “alarm” and “intelligent” terminology indicates integration of autonomous decision-making algorithms enabling predictive maintenance and real-time threat identification in monitored environments.

**Table 3** Cluster naming results

| Cluster | Cluster Name | Representative Keywords |
|---|---|---|
| 0 | Smart Monitoring and Sensor Systems | Module, Monitor, Sensor, Detection, Intelligent, Alarm, Equipment |
| 1 | Network Communication and Data Transmission | Data, Network, Communication, Signal, Terminal, Message, Receive |
| 2 | Connected Vehicle and Smart Transportation | Smart, Car, Transmission, Service, Security, Generation |
| 3 | Power Management and Energy Control Systems | Power, Energy, Circuit, Electric, Control, Connect |
| 4 | Mechanical Assembly and Hardware Architecture | Plate, Body, Arrange, Box, Rod, Mount, Shell |
| 5 | Agricultural IoT and Environmental Monitoring | Agricultural, Water, Plant, Greenhouse, Irrigation, Environment |
| 6 | Cybersecurity and Authentication Protocols | Security, Authentication, Key, Access, User, Information |
| 7 | Data Management and Intelligent Platform Systems | Platform, Management, Process, Equipment, Service, Intelligent |
| 8 | Water Resource Monitoring and Management | Meter, Pipe, Intelligent, Arrangement, Sensor, Monitor |

Cluster 1 encompasses the Network Communication and Data Transmission domain, representing the essential connectivity layer enabling device interoperability within IoT ecosystems. The dominant keywords “data,” “network,” and “communication” establish the core technological foundation for reliable information exchange across heterogeneous device populations. Patents within this cluster address fundamental communication protocol optimization, signal integrity preservation, and multi-terminal coordination mechanisms. The presence of “terminal” and “receive” keywords indicates systematic approaches to standardizing communication endpoints and data reception protocols, essential for achieving interoperability across diverse IoT implementations.

Cluster 2, designated as Connected Vehicle and Smart Transportation, represents the rapidly expanding automotive IoT domain. The distinctive keyword “smart” coupled with “car,” “transmission,” and “security” characterizes technologies enabling vehicle-to-infrastructure connectivity and autonomous transportation systems. Patents in this cluster encompass vehicle diagnostics, real-time telematics, and security protocols protecting vehicular networks from unauthorized access. The keyword “service” reflects the commercial dimension of connected vehicle technologies, addressing value-added services including predictive maintenance and driver assistance systems. This cluster represents a critical frontier in IoT commercialization and represents significant investment in intelligent transportation systems development.

Cluster 3, Power Management and Energy Control Systems, addresses sustainable infrastructure provisioning for distributed IoT networks. Keywords including “power,” “energy,” “circuit,” and “electric” characterize technologies for efficient energy consumption optimization and thermal management in resource-constrained environments. The prominence of “control” terminology indicates active power regulation and demand-responsive management mechanisms. Patents within this cluster address battery optimization techniques, wireless power transmission protocols, and energy harvesting approaches essential for enabling perpetual operation of spatially dispersed IoT devices. The technological focus on energy efficiency directly supports the scalability and sustainability requirements of large-scale IoT deployments.

Cluster 4, termed Mechanical Assembly and Hardware Architecture, represents the physical embodiment layer of IoT systems. Keywords “plate,” “body,” “arrange,” “box,” “rod,” “mount,” and “shell” collectively characterize patents focused on miniaturization, structural integration, and protective enclosure design. Although technologically distinct from software and communication layers, this domain is indispensable for IoT practical deployment, addressing industrial design innovations that enable consumer accessibility and field operability. Patents in this cluster typically document mechanical integration strategies optimizing form factor, durability, and environmental resistance for diverse operational contexts.

Cluster 5, Agricultural IoT and Environmental Monitoring, represents the application of IoT sensing technologies to precision agriculture and sustainable resource management. The distinctive keywords “agricultural,” “water,” “plant,” “greenhouse,” and “irrigation” establish the vertical domain focus on agricultural productivity optimization. Patents consolidate innovations in soil health monitoring, climate-controlled cultivation systems, and automated irrigation control mechanisms. The keyword “environment” reflects holistic environmental parameter monitoring enabling optimization of growing conditions. This cluster demonstrates IoT’s significant impact on sustainable food production, addressing critical challenges in resource optimization and agricultural yield maximization through sensor-based environmental management.

Cluster 6, designated as Cybersecurity and Authentication Protocols, delineates essential security frameworks protecting IoT ecosystem integrity. Keywords "security," "authentication," "key," and "access" establish the technological focus on cryptographic mechanisms, identity verification, and authorization frameworks [65]. Patents in this cluster address multi-factor authentication, encrypted communication protocols, and key management infrastructure. The keywords "user" and "information" indicate human-in-the-loop security considerations and information protection mechanisms. Given the inherent vulnerability of distributed IoT networks to cyber threats, this cluster represents critical enabling technologies for establishing trustworthy device-to-device and device-to-infrastructure communication, including emerging blockchain-based authentication approaches.

Cluster 7, Data Management and Intelligent Platform Systems, encompasses higher-level services enabling data aggregation, analytics, and autonomous decision-making within IoT infrastructures. Keywords "platform," "management," "process," and "intelligent" characterize middleware solutions and machine learning-based analysis systems. Patents in this cluster address cloud-IoT integration, edge computing architectures, and business intelligence extraction from heterogeneous sensor data streams. The keyword "service" indicates the commercialization of IoT analytics capabilities. This cluster represents the convergence of big data analytics, artificial intelligence, and IoT infrastructure, enabling organizations to derive actionable insights and optimize complex operational systems through data-driven methodologies.

Cluster 8, Water Resource Monitoring and Management, addresses critical infrastructure modernization through smart metering and water conservation technologies. Keywords "meter," "pipe," "arrangement," and "sensor" establish the technological focus on water distribution system monitoring and optimization. The presence of "intelligent" terminology indicates integration of machine learning algorithms for leak detection, consumption forecasting, and predictive maintenance of water infrastructure. Patents in this cluster reflect global imperatives for addressing water scarcity through IoT-enabled demand-side management and real-time system diagnostics. The technological innovations documented in this cluster directly support sustainable resource management and infrastructure resilience in water distribution networks.

The nine-cluster taxonomy identified through systematic text mining analysis establishes a comprehensive framework for understanding IoT innovation landscapes. Clusters 0, 1, 3, and 7 constitute foundational IoT layers encompassing sensing, communication, power provisioning, and data management. Clusters 2, 5, and 8 represent domain-specific IoT applications in transportation, agriculture, and water management respectively. Clusters 4 and 6 provide indispensable supporting technologies in hardware design and cybersecurity. This hierarchical taxonomy reflects the multi-disciplinary nature of IoT development, demonstrating the necessity for integrated technological approaches spanning hardware architecture, network infrastructure, application development, and security governance. The clustering results validate the heterogeneous character of patent innovations within the IoT domain, indicating sustained technological advancement across complementary technological trajectories.

### 4.4 Technology life cycle analysis

The systematic analysis of patent data through logistic growth regression modeling reveals the technological maturity trajectories and future development prospects of nine distinct IoT innovation clusters. Table 4 synthesizes the regression analysis results, quantifying each cluster's progression through technology life cycle stages and projecting future market saturation dynamics based on fitted S-curve parameters. The analysis incorporates comprehensive patent filing data spanning over five decades (1969–2024), with year 2025 excluded from modeling to ensure temporal validity and predictive accuracy.

#### *4.4.1 Model validation and methodological robustness*

The logistic regression models demonstrate exceptional explanatory power, with $R^2$ values ranging from 0.9987 to 0.9998, indicating that the S-curve specification captures with remarkable precision the underlying dynamics of patent emergence, acceleration, and diffusion within each technological domain [62]. These exceptionally high goodness-of-fit coefficients substantially exceed conventional thresholds for technology forecasting models, validating both the appropriateness of the logistic growth framework and the systematic nature of innovation diffusion in IoT technologies [50].The outstanding model performance across all clusters—with the lowest $R^2$ value at 0.9987 for Power Management and Energy Control Systems and Mechanical Assembly and Hardware Architecture—demonstrates that IoT patent filing patterns follow highly predictable S-shaped trajectories consistent with established technology diffusion theories [39, 136]. The near-perfect fit for Cybersecurity and Authentication Protocols ($R^2$ = 0.9998) and Data Management and Intelligent Platform Systems ($R^2$ = 0.9998) reflects the structured, standards-driven evolution characteristic of foundational digital infrastructure technologies [34].

The fitted parameters provide critical insights into technology diffusion dynamics. The carrying capacity (k) values reveal substantial variation in ultimate market potential, ranging from 5,203 patents for Connected Vehicle and Smart Transportation to 42,001 patents for Data Management and Intelligent Platform Systems. This 8-fold differential reflects both the breadth of technological scope and the diversity of application contexts within each cluster [22]. The inflection points (a) demonstrate remarkable temporal convergence, clustering tightly between 2020.60 and 2021.95, indicating synchronized acceleration of innovation activity across disparate IoT domains during the 2020–2022 period—likely catalyzed by pandemic-driven digital transformation imperatives and 5G infrastructure deployment [98]. The steepness parameters (b) ranging from 1.06 to 2.45 quantify the velocity of technology diffusion, with lower values indicating rapid market penetration and higher values suggesting more gradual adoption trajectories.

#### *4.4.2 Historical development patterns and Temporal dynamics*

Analysis of emergence dates reveals a multi-decadal innovation trajectory, with the earliest cluster—Network Communication and Data Transmission—commencing patent activity in 1969, coinciding with the foundational period of packet-switching networks and digital telecommunications infrastructure [156]. This 55-year development arc underscores that contemporary IoT communication technologies build upon decades of incremental innovation in network protocols, wireless transmission standards, and data encoding mechanisms [4]. Similarly, Data Management and Intelligent Platform Systems demonstrates patent activity originating in 1978, reflecting the evolution from early database management systems and distributed computing architectures through contemporary cloud-IoT integration platforms and edge analytics frameworks [69, 104]. The 46-year development trajectory highlights the cumulative nature of software platform innovation, where contemporary intelligent systems integrate decades of advances in data modeling, query optimization, distributed processing, and machine learning algorithms.

**Table 4** Patent technology life cycle stage division with regression model performance

| Cluster Number | Cluster Name | Emerging | Growth | Maturity | Saturation | $R^2$ | k | a | b |
|---|---|---|---|---|---|---|---|---|---|
| 0 | Smart Monitoring and Sensor Systems | **1990** | **2017** | **2022** | **2027** | **0.9996** | **29893.22** | **2021.68** | **2.29** |
| 1 | Network Communication and Data Transmission | **1969** | **2017** | **2021** | **2025** | **0.9994** | **23002.62** | **2021.01** | **1.92** |
| 2 | Connected Vehicle and Smart Transportation | **2015** | **2018** | **2021** | **2024** | **0.9993** | **5203.32** | **2020.86** | **1.38** |
| 3 | Power Management and Energy Control Systems | **1996** | **2016** | **2021** | **2026** | **0.9987** | **21851.64** | **2020.87** | **2.11** |
| 4 | Mechanical Assembly and Hardware Architecture | **1991** | **2018** | **2021** | **2023** | **0.9987** | **24076.14** | **2020.69** | **1.06** |
| 5 | Agricultural IoT and Environmental Monitoring | **2002** | **2016** | **2021** | **2025** | **0.9994** | **7731.21** | **2020.60** | **2.02** |
| 6 | Cybersecurity and Authentication Protocols | **2008** | **2017** | **2021** | **2026** | **0.9998** | **13213.90** | **2021.39** | **1.92** |
| 7 | Data Management and Intelligent Platform Systems | **1978** | **2017** | **2022** | **2027** | **0.9998** | **42000.64** | **2021.95** | **2.45** |
| 8 | Water Resource Monitoring and Management | **1984** | **2017** | **2021** | **2025** | **0.9991** | **11333.76** | **2020.82** | **1.68** |

Water Resource Monitoring and Management (emergence: 1984) and Mechanical Assembly and Hardware Architecture (emergence: 1991) represent intermediate-timeline clusters with 40-year and 33-year development histories, respectively. These domains reflect sustained innovation in physical sensing infrastructure, embedded systems integration, and industrial automation—technologies requiring extended refinement cycles due to reliability requirements, environmental constraints, and integration with legacy infrastructure [73, 136].

In contrast, more recent emergence dates characterize application-specific domains: Cybersecurity and Authentication Protocols (2008), Agricultural IoT and Environmental Monitoring (2002), and Connected Vehicle and Smart Transportation (2015). These clusters represent specialized applications enabled by maturation of foundational technologies, demonstrating the hierarchical nature of IoT innovation where platform-level capabilities precede domain-specific implementations [42, 136].

As illustrated in Fig. 7, the S-curve visualizations demonstrate distinct acceleration phases, with all clusters exhibiting pronounced upward inflection during the 2015–2020 period, followed by deceleration toward saturation as patent filing rates stabilize. This synchronized acceleration reflects ecosystem-wide technological convergence, standardization of interoperability protocols, and commercialization of integrated IoT solutions [110, 111, 136].

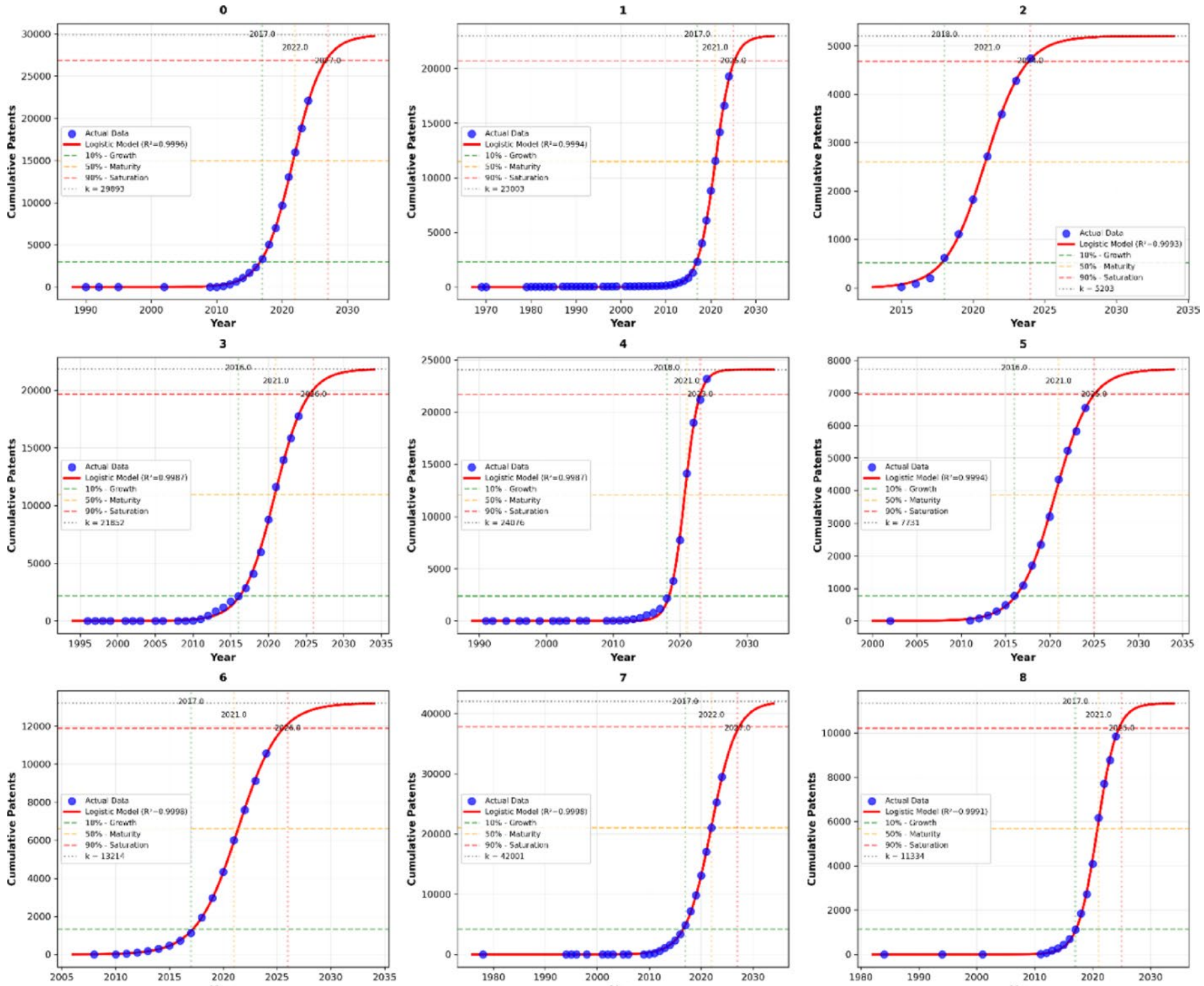

**Fig. 7** S-Curve Technology Life Cycle Trajectories for IoT Patent Clusters

#### *4.4.3 Current technology maturity status*

**4.4.3.1** Technologies approaching Near-Term saturation (2023–2025) Mechanical Assembly and Hardware Architecture represents the most advanced cluster in the maturity lifecycle, reaching projected saturation in 2023. Despite a relatively early emergence date (1991), the technology demonstrated an extended latency period before entering rapid growth in 2018. The exceptionally low steepness parameter (b = 1.06) indicates the most rapid diffusion trajectory among all clusters, reflecting accelerated standardization of modular hardware architectures, miniaturized form factors, and integration-ready component designs [30]. The high $R^2$ (0.9987) validates this rapid saturation trajectory, suggesting that fundamental hardware architecture innovations have achieved market maturity, with future advances concentrated on incremental optimization and cost reduction rather than architectural transformation.

Connected Vehicle and Smart Transportation achieves saturation in 2024 following the shortest development cycle among all clusters—spanning only nine years from emergence (2015) to saturation. This accelerated trajectory reflects intensive R&D investment driven by automotive industry transformation, regulatory mandates for connected safety systems, and strategic positioning for autonomous vehicle deployment [81]. The moderate steepness parameter (b = 1.38) indicates relatively rapid diffusion, while the small carrying capacity (k = 5,203.32) suggests a more specialized technological domain with narrower scope compared to foundational infrastructure clusters. The visualization in Fig. 1 demonstrates a steep ascent phase between 2018 and 2021, followed by deceleration toward saturation—consistent with market consolidation around established vehicular communication standards (DSRC/C-V2X) and connected telematics platforms [82].

Network Communication and Data Transmission and Water Resource Monitoring and Management both reach saturation in 2025, though via markedly different developmental pathways. Network Communication demonstrates the longest historical trajectory (56 years from emergence to saturation), reflecting continuous evolution from early telecommunication protocols through contemporary 5G-enabled IoT connectivity [94]. The inflection point at a = 2021.01 and steepness parameter b = 1.92 indicate that despite decades of incremental innovation, the most intensive patent acceleration occurred recently, driven by Low Power Wide Area Network (LPWAN) protocols, mesh networking architectures, and millimeter-wave communication technologies [108].

Conversely, Water Resource Monitoring exhibits a 41-year development arc with substantially smaller carrying capacity (k = 11,333.76), reflecting a specialized application domain addressing water distribution infrastructure, consumption analytics, and leak detection systems [54].The convergence toward 2025 saturation, as shown in Fig. 1, indicates consolidation of sensor deployment methodologies, wireless transmission protocols for underground infrastructure, and data analytics frameworks for consumption forecasting [114].

Agricultural IoT and Environmental Monitoring reaches saturation in 2025 following a 23-year development cycle. Despite earlier assumptions of heterogeneous development patterns and lower predictive accuracy in agricultural technologies, the exceptionally high $R^2$ (0.9994) demonstrates that precision agriculture innovations follow systematic diffusion trajectories [136]. The inflection point at a = 2020.60 indicates that agricultural IoT experienced maximum acceleration during 2020–2021, potentially catalyzed by supply chain disruptions, labor shortages, and increased emphasis on agricultural resilience during the COVID-19 pandemic [156].

The moderate steepness (b = 2.02) reflects ongoing integration of diverse sensor modalities (soil moisture, atmospheric conditions, crop health imaging) into unified farm management platforms [83].

**4.4.3.2** Technologies reaching Medium-Term saturation (2026–2027) Power Management and Energy Control Systems approaches saturation in 2026 after a 30-year development trajectory from emergence in 1996. The substantial carrying capacity (k = 21,851.64) reflects the breadth of innovations encompassing energy harvesting, wireless power transfer, battery optimization, and dynamic power allocation algorithms [136]. The inflection point at a = 2020.87 indicates peak innovation velocity during 2020–2021, driven by ultra-low-power system-on-chip architectures and ambient energy harvesting technologies enabling battery-free IoT deployments [100].

As depicted in Fig. 1, the technology demonstrates steady progression through maturity phase (2021–2026) with deceleration indicating consolidation around dominant design paradigms for energy-constrained devices.

Cybersecurity and Authentication Protocols reaches saturation in 2026 despite relatively recent emergence in 2008, demonstrating the most rapid growth trajectory among mature infrastructure technologies. The exceptional model fit ($R^2$ = 0.9998) and inflection point at a = 2021.39 indicate that IoT security innovations follow highly predictable, standards-driven diffusion patterns [61]. The moderate steepness (b = 1.92) and substantial carrying capacity (k = 13,213.90) reflect intensive patent activity addressing cryptographic frameworks, blockchain-based authentication, physical unclonable functions (PUFs), and quantum-resistant security protocols [6, 56]. The near-saturation status suggests that foundational security architectures have matured, though the domain continues evolving through integration of zero-trust models and confidential computing frameworks [74].

Smart Monitoring and Sensor Systems and Data Management and Intelligent Platform Systems both achieve saturation in 2027, representing foundational infrastructure technologies with the largest carrying capacities (k = 29,893.22 and k = 42,000.64, respectively). Smart Monitoring demonstrates a 37-year development arc with inflection point a = 2021.68 and steepness b = 2.29, reflecting sustained innovation across diverse sensing modalities, edge-based signal processing, and multi-sensor fusion architectures [50, 136]. The visualization in Fig. 1 shows continued growth through 2024 with gradual deceleration toward 2027 saturation, indicating ongoing refinement of miniaturization techniques, MEMS sensor integration, and energy-efficient sensing protocols.

Data Management and Intelligent Platform Systems exhibits the largest carrying capacity among all clusters (k = 42,000.64), reflecting the scope and diversity of middleware solutions, cloud-IoT integration frameworks, edge computing architectures, and machine learning-based analytics platforms [124, 136]. The extended 49-year development timeline from emergence (1978) through projected saturation (2027) demonstrates the evolutionary trajectory from early database systems through contemporary intelligent platforms incorporating real-time stream processing, federated learning, and AI-driven automation [112, 133].

#### *4.4.4 Temporal convergence and ecosystem synchronization*

A striking finding evident in Table 1; Fig. 1 is the remarkable temporal convergence of saturation dates across technologically diverse clusters. Despite emergence dates spanning 46 years (1969–2015), eight of nine clusters reach saturation within a compressed five-year window (2023–2027), with only one cluster (Mechanical Assembly) achieving saturation by 2023. This convergence suggests ecosystem-wide synchronization of technology maturation dynamics, driven by:

1. *Standardization Imperatives* The proliferation of interoperability standards (MQTT, CoAP, OPC UA) and certification frameworks has accelerated adoption by reducing integration barriers and enabling plug-and-play deployment models [96, 115, 136].
2. *Platform Consolidation* Emergence of dominant cloud-IoT platforms (AWS IoT, Azure IoT, Google Cloud IoT) has created standardized deployment pathways that accelerate technology diffusion across application domains [51, 137].
3. *5G Infrastructure Deployment* Widespread 5G network rollout has eliminated connectivity constraints that previously limited IoT scalability, enabling synchronized advancement across communication-dependent technologies [36, 136].
4. *Pandemic-Driven Digital Acceleration* The COVID-19 pandemic catalyzed accelerated adoption of remote monitoring, automation, and contactless systems, creating synchronized demand across diverse IoT domains [31, 136].

The clustering of inflection points (a parameters) between 2020.60 and 2021.95 provides quantitative evidence of this synchronized acceleration, with maximum growth rates achieved during 2020–2022 across all technology clusters. This temporal alignment transcends individual technology characteristics, suggesting external market forces and ecosystem-level dynamics dominate individual cluster development trajectories [16].

### 4.5 IoT patents registration network analysis

#### *4.5.1 Network structure and characteristics*

The IoT patent collaboration network provides critical insights into the structural dynamics of innovation partnerships within the Internet of Things domain. This network was constructed based on co-registration relationships, wherein two or more entities jointly file patent applications. Each node in the network represents a distinct entity—either an individual inventor or an organizational assignee—that has participated in patent registration activities. The edges connecting nodes signify collaborative relationships, with edge weight representing the frequency of joint patent filings between connected entities. The network was visualized and analyzed using Gephi software, a widely adopted platform for network analysis and visualization [17]. The resulting network structure is illustrated in Fig. 8, which depicts the complex web of collaborative relationships among IoT innovators.

The fundamental structural characteristics of the IoT patent collaboration network are presented in Table 5. The network comprises 16,845 nodes and 34,404 edges, revealing a substantial innovation ecosystem with diverse participants. The average degree of 4.085 indicates that each entity typically engages in collaborative relationships with approximately four other actors within the network. The average weighted degree of 11.771, which is considerably higher than the simple average degree, suggests that certain collaborative partnerships are particularly productive, generating multiple joint patent applications [116]. This disparity between simple and weighted degree metrics indicates heterogeneity in collaboration intensity across the network.

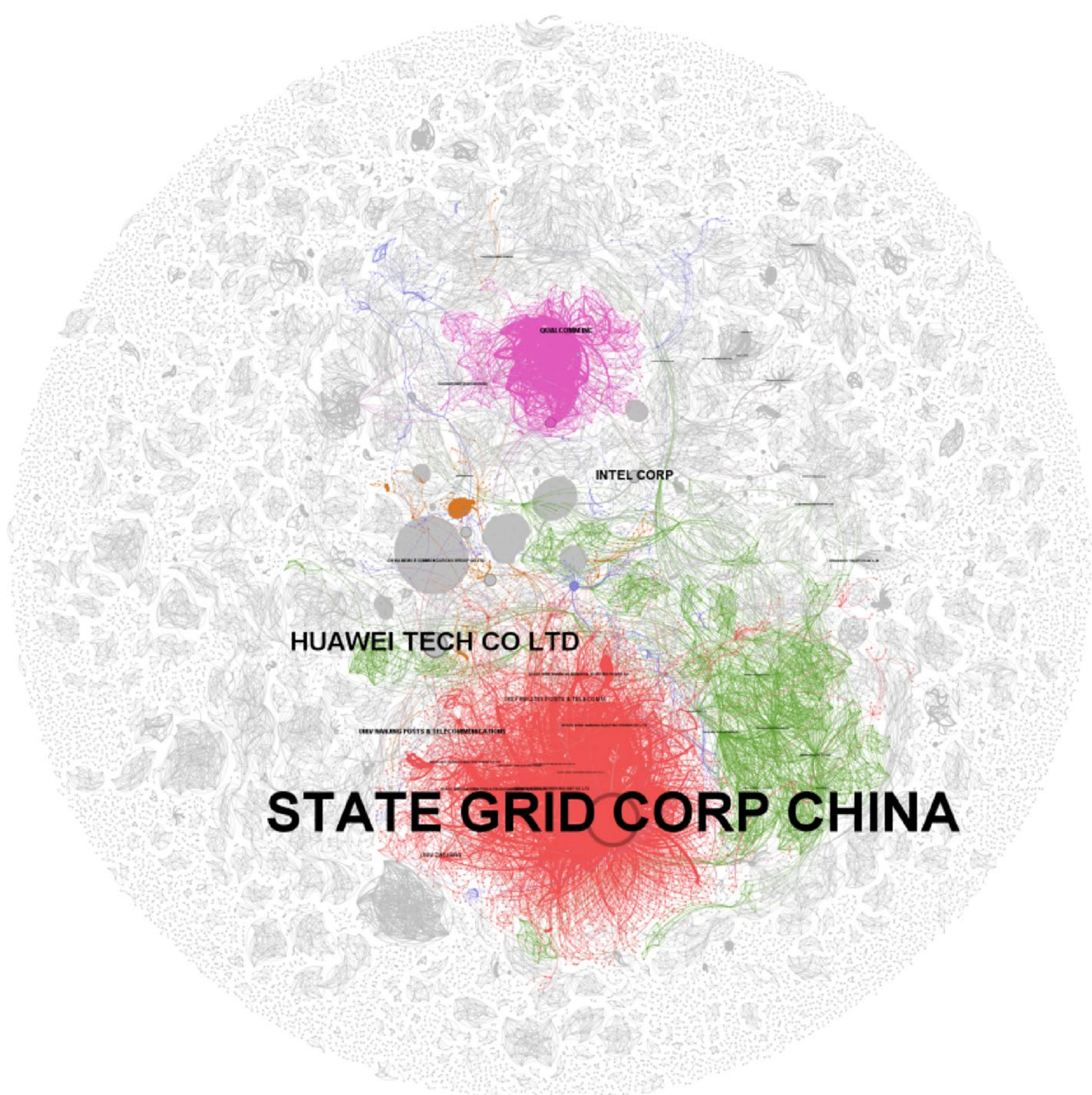

**Fig. 8** IoT patent collaboration network visualization

**Table 5** Key metrics of the IoT patent collaboration network

| Metric | Value |
|---|---|
| Nodes | 16,845 |
| Edges | 34,404 |
| Average Degree | 4.085 |
| Avg. Weighted Degree | 11.771 |
| Network Diameter | 21 |
| Graph Density | 0.001 |
| Modularity | 0.963 |
| Avg. Clustering Coefficient | 0.855 |
| Avg. Path Length | 5.754 |

The network density of 0.001 indicates a sparse connectivity structure, wherein only 0.1% of all possible connections are realized. This sparsity suggests that collaborative relationships in IoT patent development are characterized by selective partnerships rather than diffuse interconnectedness across the entire field. Such selectivity may reflect specialization patterns, technological proximity requirements, or strategic alliance formation based on complementary capabilities [71].

The average clustering coefficient of 0.855 demonstrates a pronounced tendency for nodes to form tightly-knit collaborative clusters, wherein the partners of a given entity are also likely to collaborate with one another. This high clustering indicates the presence of cohesive research communities or institutional partnerships within the broader IoT innovation landscape [156]. Concurrently, the average path length of 5.754 and network diameter of 21 reveal that despite the sparse overall connectivity, the network exhibits relatively short distances between nodes. This combination of high clustering and short path lengths is characteristic of "small-world" network topology, which facilitates efficient knowledge diffusion within local communities while maintaining global connectivity through strategic bridging relationships [118].

The exceptionally high modularity value of 0.963 provides strong evidence for the existence of distinct communities within the network, each potentially concentrating on specific technological subdomains or application areas within IoT innovation. Modularity measures the degree to which a network can be partitioned into densely connected modules with sparse connections between modules [119]. The visualization in Fig. 8 employs distinct color coding to represent these communities, where each color corresponds to a community comprising more than 1% of the total network participants. This community structure reflects the multifaceted nature of IoT technology development and the tendency for specialized clusters to emerge around particular technological challenges or application domains [136].

#### *4.5.2 Centrality analysis*

To identify the most influential actors within the IoT patent collaboration network, four complementary centrality measures were computed: degree centrality, betweenness centrality, closeness centrality, and eigenvector centrality. Each measure captures different dimensions of network influence and strategic positioning. Table 6 presents the top 10 entities for each centrality metric, revealing both consistencies and divergences in network prominence across different analytical perspectives.

State Grid Corp China emerges as the overwhelmingly dominant actor across all four centrality measures, occupying the top position in degree, betweenness, closeness, and eigenvector centrality. This universal prominence indicates that State Grid functions simultaneously as the most connected entity (degree), the most critical bridge between different network regions (betweenness), the most efficiently positioned for rapid information access (closeness), and the most influential through connections to other highly connected nodes (eigenvector). This multidimensional centrality reflects State Grid's strategic role as the central orchestrator of China's smart grid IoT innovation ecosystem, coordinating diverse stakeholders including provincial utilities, research institutions, and technology providers. The degree centrality rankings reveal Indian academic researchers (Kshirsagar, Manjunath, Peroumal) among the top 10, reflecting their extensive collaborative networks within the Indian IoT Academic Research Network community, though these researchers rank lower or disappear entirely in other centrality measures, indicating that their influence remains largely confined to densely connected academic clusters rather than extending to broader network brokerage or efficiency positions (Abbasi et al., 2012). Qualcomm Inc and Intel Corp appear prominently across multiple centrality measures, reflecting their roles as major corporate innovation hubs, though their rankings vary: Qualcomm ranks second in degree centrality but eighth in betweenness, while Intel ranks eighth in degree but third in betweenness, suggesting Intel occupies more strategic bridging positions connecting disparate network regions despite fewer total partnerships.

**Table 6** Top 10 entities by centrality measures in the IoT patent collaboration network

| Rank | Degree Centrality | Betweenness Centrality | Closeness Centrality | Eigenvector Centrality |
|---|---|---|---|---|
| 1 | State Grid Corp China | State Grid Corp China | State Grid Corp China | State Grid Corp China |
| 2 | Qualcomm Inc | Huawei Tech Co Ltd | Huawei Tech Co Ltd | State Grid Information & Telecommunication Co Ltd |
| 3 | Kshirsagar Pravin R Dr | Intel Corp | China Electric Power Res Inst Co Ltd | China Electric Power Res Inst Co Ltd |
| 4 | Manjunath T C Dr | Li Yong | Univ Southeast | State Grid Jiangsu Electric Power Co Ltd |
| 5 | State Grid Information & Telecommunication Co Ltd | Li Wei | State Grid Information & Telecommunication Co Ltd | Kshirsagar Pravin R Dr |
| 6 | Peroumal Vijayakumar Dr | Yang Wei | Univ Nanjing Posts & Telecommunications | Global Energy Interconnection Res Inst Co Ltd |
| 7 | State Grid Jiangsu Electric Power Co Ltd | Univ Tsinghua | Univ Zhejiang | Manjunath T C Dr |
| 8 | Intel Corp | Qualcomm Inc | China Electric Power Res Inst | Univ Beijing Posts & Telecomm |
| 9 | China Electric Power Res Inst Co Ltd | Univ Nanjing Posts & Telecommunications | State Grid Jiangsu Electric Power Co Ltd | Qualcomm Inc |
| 10 | Univ Beijing Posts & Telecomm | Univ Beijing Posts & Telecomm | Univ Beijing Posts & Telecomm | Peroumal Vijayakumar Dr |

The betweenness centrality rankings reveal notable differences from degree-based measures, with Huawei Tech Co Ltd, Li Yong, Li Wei, Yang Wei, and Univ Tsinghua appearing prominently despite lower or absent rankings in degree centrality. This pattern indicates these actors occupy critical bridging positions connecting otherwise disconnected network components, potentially serving as knowledge brokers between different technological domains, geographic regions, or institutional types even without maintaining the highest number of direct collaborations. The prevalence of individual inventors (Li Yong, Li Wei, Yang Wei) in betweenness rankings but not degree rankings suggests these individuals may be serial collaborators working across multiple organizational contexts or technological domains, accumulating structural advantage through boundary-spanning rather than partnership volume.

The closeness centrality rankings show strong presence of Chinese universities (Univ Southeast, Univ Nanjing Posts & Telecommunications, Univ Zhejiang, Univ Beijing Posts & Telecomm) and research institutions, indicating these entities occupy positions enabling efficient information access and dissemination throughout the network, potentially serving as knowledge integrators within the Chinese IoT innovation system.

The eigenvector centrality rankings closely mirror degree centrality for top positions, with State Grid ecosystem entities (State Grid Corp China, State Grid Information & Telecommunication, State Grid Jiangsu Electric Power) dominating alongside Indian academic researchers, confirming these actors' influence derives not merely from connection quantity but from connections to other highly connected entities within densely connected communities (Bonacich, 1987). The convergence of multiple State Grid entities in eigenvector rankings reflects the tightly integrated nature of the China State Grid IoT Consortium, where provincial utilities and affiliated research institutes form a mutually reinforcing influence network.

#### *4.5.3 Community detection and characterization*

The application of the Louvain community detection algorithm revealed several significant collaborative communities within the IoT patent network. The algorithm identifies clusters of densely interconnected nodes that exhibit stronger internal connections than external linkages, thereby revealing specialized collaborative groups [29]. Communities comprising at least 1% of the total network members are highlighted with distinct colors in Fig. 8. Detailed information about these principal communities is presented in Table 7, which includes the community identifier, proposed name, and the most prominent members based on degree centrality.

The nomenclature assigned to each community was determined through expert consensus based on the organizational composition, institutional affiliations, and collaborative patterns evident within each cluster. The naming strategy emphasizes the dominant organizational characteristics and geographical or sectoral concentration of community members, following established practices in innovation network analysis [2].

The red community is designated as "China State Grid IoT Consortium" based on the overwhelming dominance of State Grid Corporation of China and its affiliated provincial subsidiaries, which constitute the core membership structure. The term "consortium" is justified by the presence of coordinated collaboration between a central state-owned enterprise and its distributed organizational units, characteristic of formal innovation consortia in infrastructure sectors [101]. The inclusion of multiple provincial electric power companies alongside research institutes represents a vertically integrated collaborative structure typical of national strategic technology initiatives in the Chinese innovation system [97].

**Table 7** Community detection analysis results

| Color | Community Name | Most Important Members |
| --- | --- | --- |
| Red | China State Grid IoT Consortium | State Grid Corp China, State Grid Information & Telecommunication Co. Ltd, State Grid Jiangsu Electric Power Co. Ltd, China Electric Power Res Inst Co. Ltd, Univ Beijing Posts & Telecomm, State Grid Zhejiang Electric Power Co. Ltd, Univ Zhejiang, Univ North China Electric Power, State Grid Shanghai Municipal Electric Power Co, Global Energy Interconnection Res Inst Co. Ltd |
| Green | Indian IoT Academic Research Network | Kshirsagar Pravin R Dr, Manjunath T C Dr, Peroumal Vijayakumar Dr, G Pavithra Dr, Yaseen Syed Mufassir Mr, Dadheech Pankaj Dr, Kumar Chinnapalli Likith Mr, Gulati Kamal Dr, Indra Gaurav, Kumar Abhishek Dr |
| Blue | Chinese Hydropower & Clean Energy Research Alliance | Univ Tsinghua, Univ Hohai, Xian Thermal Power Res Inst Co, Nat Computer Network & Inf Security Management Ct, Huaneng Clean Energy Res Inst, Powerchina Huadong Engineering Corp Ltd, Sinohydro Bureau 3 Co Ltd, Sinohydro Bureau 14 Co Ltd, Huaneng Lancang River Hydropower Co Ltd, China Railway Design Corp |
| Pink | Global Telecommunications Technology Leaders | Qualcomm Inc, Wu Zhikun, Kim Yuchul, Wei Chao, Chen Wanshi, Gaal Peter, Ji Tingfang, Gupta Piyush, Yang Wei, Phuyal Umesh, Liu Le, He Linhai, Elshafie Ahmed |
| Orange | Chinese Smart Technology & University Consortium | Midea Group Co Ltd, Univ Electronic Sci & Tech China, Cernet Co Ltd, Univ Guangxi, Univ Jinan, Univ Dalian Tech, Electric Power Res Inst Co Ltd CSG, China Southern Power Grid Co, Univ Chongqing Posts & Telecom, Univ Nantong, Univ Tianjin Science & Tech, Shenzhen Power Supply Bureau |

The green community is labeled "Indian IoT Academic Research Network" due to the exclusive composition of individual academic researchers, predominantly identified through doctoral titles and Indian institutional affiliations. This nomenclature reflects the network's academic character and geographical concentration, distinguishing it from corporate or mixed-sector communities [156]. The term "network" appropriately captures the decentralized, peer-to-peer collaborative structure characteristic of academic research communities, as opposed to hierarchical organizational arrangements [84].

The blue community is named "Chinese Hydropower & Clean Energy Research Alliance" based on the predominance of hydropower engineering enterprises (Sinohydro bureaus, Huaneng entities) and leading universities specializing in hydraulic engineering (Tsinghua University, Hohai University). The designation "alliance" reflects the strategic partnership between construction enterprises, energy companies, and research institutions focused on a specific technological domain [72]. The sectoral specificity toward hydropower and clean energy distinguishes this community from broader energy-focused clusters [122].

The pink community is designated "Global Telecommunications Technology Leaders" reflecting the anchoring presence of Qualcomm Inc., a multinational telecommunications technology corporation, and the international diversity of co-inventors from multiple continents. The term "global" emphasizes the geographically distributed composition, distinguishing this community from regionally concentrated clusters [18]. The designation "leaders" acknowledges Qualcomm's established position in wireless communication standards and IoT connectivity technologies [35].

The orange community is labeled "Chinese Smart Technology & University Consortium" based on the hybrid composition combining Midea Group Co Ltd—a major consumer electronics and appliance manufacturer—with multiple geographically distributed Chinese universities. This nomenclature captures the industry-academia collaborative structure characteristic of Chinese innovation ecosystems [46]. The term "smart technology" reflects Midea's focus on intelligent consumer products, while "consortium" denotes the coordinated multi-institutional structure bridging commercial and academic sectors [21].

## 5 Discussion

The cross-analysis of IoT technology clusters and patent collaboration communities reveals distinct innovation strategies and technological specialization patterns that illuminate the heterogeneous nature of IoT development. By examining the intersection of technological domains with collaborative networks, we can discern how different communities prioritize specific technological areas based on their organizational capabilities, sectoral requirements, and strategic positioning. Table 8 presents the percentage distribution of IoT technology clusters across the identified communities, while Fig. 9 illustrates these patterns through a Sankey diagram that visualizes the flow of innovation activities from communities to technology clusters.

The analysis reveals fundamentally distinct innovation orientations among the identified communities, reflecting their divergent organizational mandates, sectoral contexts, and technological capabilities. The China State Grid IoT Consortium exhibits a diversified technology portfolio with concentration in Data Management and Intelligent Platform Systems (22.86%), Power Management and Energy Control Systems (20.31%), and Smart Monitoring and Sensor Systems (17.74%), reflecting the infrastructure-centric mandate of state grid operations requiring comprehensive smart grid capabilities [7]. The substantial investment in cybersecurity (13.71%) and network communication (12.15%) underscores the critical importance of secure, reliable connectivity for critical infrastructure operations [136].

**Table 8** Percentage distribution of IoT technology clusters across key communities

| Community | Smart Monitoring and Sensor Systems | Network Communication and Data Transmission | Connected Vehicle and Smart Transportation | Power Management and Energy Control Systems | Mechanical Assembly and Hardware Architecture | Agricultural IoT and Environmental Monitoring | Cybersecurity and Authentication Protocols | Data Management and Intelligent Platform Systems | Water Resource Monitoring and Management |
|---|---|---|---|---|---|---|---|---|---|
| China State Grid IoT Consortium | 17.7% | 12.2% | 0.3% | 20.3% | 6.0% | 3.5% | 13.7% | 22.9% | 3.5% |
| Chinese Hydropower & Clean Energy Research Alliance | 18.5% | 8.3% | 0.1% | 10.5% | 7.8% | 8.3% | 12.6% | 24.0% | 9.9% |
| Chinese Smart Technology & University Consortium | 15.6% | 12.1% | 0.1% | 21.0% | 3.2% | 4.1% | 13.3% | 26.8% | 3.9% |
| Global Telecommunications Technology Leaders | 1.8% | 87.7% | 0.1% | 2.1% | 4.5% | 0.2% | 2.5% | 0.8% | 0.4% |
| Indian IoT Academic Research Network | 15.1% | 13.3% | 9.6% | 3.0% | 1.8% | 18.1% | 7.8% | 24.7% | 6.6% |

The Chinese Hydropower & Clean Energy Research Alliance demonstrates similar platform emphasis (23.95%) and monitoring systems focus (18.49%), but differs significantly in its higher investment in water resource monitoring (9.94%) and agricultural IoT (8.26%), reflecting the integrated nature of hydropower operations that must simultaneously address hydrological monitoring, environmental compliance, and watershed management [80]. The lower power management investment (10.50%) compared to the State Grid Consortium suggests different operational priorities between generation facilities and distribution networks.

The Chinese Smart Technology & University Consortium exhibits the highest platform concentration (26.78%), reflecting platform-centric business models where ecosystem control and data analytics constitute core competitive advantages [45]. The substantial power management investment (21.00%) aligns with smart home applications emphasizing energy efficiency, while lower mechanical assembly focus (3.23%) indicates reliance on contract manufacturing rather than proprietary hardware development [110, 111].

The Global Telecommunications Technology Leaders community demonstrates extreme specialization with 87.70% concentration in Network Communication and Data Transmission, reflecting focused expertise in connectivity infrastructure that enables diverse IoT applications [125]. This concentration strategy appears rational given communication technologies' critical bottleneck role in IoT ecosystem development [4]. The minimal investment across application domains (0.17–2.53%) indicates clear strategic boundaries, providing enabling infrastructure while leaving application development to domain specialists.

The Indian IoT Academic Research Network presents a distinctive profile with notable emphasis on Agricultural IoT and Environmental Monitoring (18.07%), substantially exceeding all corporate communities and suggesting research alignment with India's agricultural sector challenges and rural development priorities [55]. The balanced platform investment (24.70%) combined with relatively high connected vehicle focus (9.64%)—contrasting sharply with other communities (0.07–0.26%)—indicates academic research addressing transportation challenges and exploring emerging applications unconstrained by commercial pressures [85]. The lower power management (3.01%) and mechanical assembly (1.81%) investments reflect academic emphasis on software, algorithms, and system architectures rather than hardware engineering.

These divergent patterns reveal fundamental complementarities within the IoT innovation ecosystem. Infrastructure operators pursue portfolio diversification addressing multifaceted smart infrastructure requirements; consumer technology companies prioritize platforms and power management for ecosystem control; telecommunications specialists maintain focused investment in connectivity infrastructure; and academic researchers pursue development-oriented research agendas. This division of innovative labor suggests vertical specialization where different communities contribute distinct technological capabilities collectively enabling comprehensive IoT solutions [63].

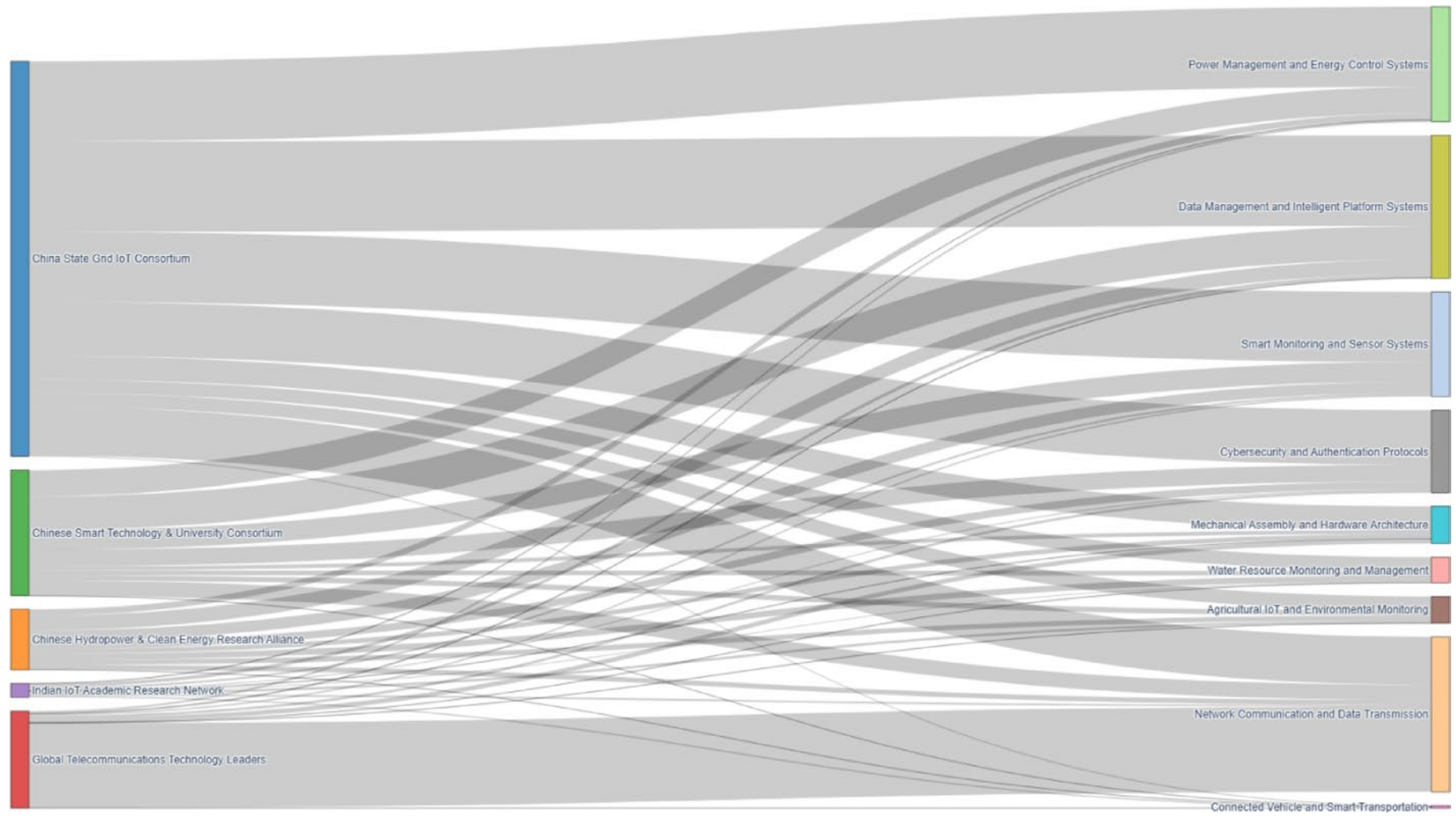

**Fig. 9** Patent distribution of communities across identified technology clusters

The uniformly low connected vehicle investment across corporate communities (0.07–0.26%) except academics (9.64%) suggests either distinct automotive innovation networks not captured in identified communities, or sectoral boundaries limiting IoT infrastructure providers' automotive engagement [64]. The Chinese communities' balanced portfolios compared to global telecommunications' extreme focus may reflect institutional characteristics of China's innovation system emphasizing comprehensive technological development versus market-driven specialization [156]. The consistent high cybersecurity investment across Chinese communities (12.61–13.71%) reflects both strategic importance assigned to information security and regulatory requirements.

### 5.1 Managerial insights and implementation

The integrated analysis of technology clusters and collaboration networks offers actionable intelligence for diverse stakeholders navigating the IoT innovation landscape. For technology corporations, the identification of near-saturation domains such as Mechanical Assembly and Hardware Architecture (saturation: 2023) and Connected Vehicle Technologies (saturation: 2024) signals diminishing returns from incremental innovation investments in these areas. Strategic emphasis should pivot toward emerging or growth-stage clusters where competitive differentiation remains achievable. Conversely, the convergence of inflection points across multiple clusters during 2020–2022 demonstrates ecosystem-wide synchronization effects, suggesting that platform-level capabilities and interoperability standards have become critical competitive differentiators rather than domain-specific technical superiority.

Small and medium enterprises seeking market entry face distinct strategic imperatives based on their resource constraints and organizational capabilities. The extreme specialization exhibited by the Global Telecommunications Technology Leaders community—allocating 87.7% of innovation efforts to network communication—validates focused positioning strategies that leverage deep expertise in enabling infrastructure rather than pursuing broad portfolio diversification. Academic institutions and research organizations can derive guidance from the Indian IoT Academic Research Network's emphasis on Agricultural IoT and Environmental Monitoring (18.07%), which substantially exceeds corporate investment levels and demonstrates alignment between research agendas and regional development priorities. This pattern suggests opportunities for academic-industry partnerships addressing underserved application domains where commercial incentives remain insufficient to drive private sector innovation.

Policymakers confronting decisions regarding research funding allocation, infrastructure investment, and regulatory frameworks can leverage life cycle classifications to optimize intervention timing. Technologies approaching saturation—particularly those with exceptionally high $R^2$ values indicating predictable trajectories—require governance frameworks emphasizing standardization, interoperability mandates, and security certification rather than exploratory research funding. Conversely, growth-stage technologies such as Data Management and Intelligent Platform Systems (saturation: 2027, $k = 42{,}001$) represent domains where strategic public investment in foundational research, workforce development initiatives, and demonstration projects can accelerate diffusion and strengthen national competitive positioning. The pronounced concentration of Chinese communities across multiple high-investment clusters underscores the effectiveness of coordinated industry-academia-government consortia in mobilizing complementary resources, suggesting institutional models worthy of adaptation in other national contexts.

### 5.2 Limitations and future research directions

This study acknowledges several methodological limitations that constrain interpretive scope and suggest directions for future inquiry. The reliance on Lens.org as the sole patent data source, while justifiable for accessibility and replicability considerations, introduces potential coverage gaps relative to specialized commercial databases. Future research should conduct comparative analyses incorporating multiple patent repositories including the World Intellectual Property Organization (WIPO) Global Brand Database and direct United States Patent and Trademark Office (USPTO) datasets to validate cluster stability and assess whether data source selection influences community detection outcomes or technology life cycle classifications.

The clustering methodology employed k-means with Davies-Bouldin Index optimization, yet alternative approaches including Silhouette analysis, hierarchical agglomerative clustering, or density-based methods may yield different technological taxonomies warranting comparative evaluation. The selection of BERT-based patent embeddings over simpler term frequency-inverse document frequency representations or Word2Vec alternatives was predicated on transformer architectures' demonstrated superiority in capturing semantic relationships, but systematic ablation studies quantifying performance gains relative to computational costs remain absent. Future methodological research should conduct controlled comparisons assessing whether BERT embeddings introduce overfitting risks or whether simpler models achieve comparable clustering validity at reduced computational expense.

Critically, this study establishes associational patterns between community membership and technological specialization but does not examine statistical or causal relationships between network dynamics and technology emergence timing. The convergence of inflection points during 2020–2022 suggests ecosystem-wide synchronization effects, yet the underlying mechanisms—whether driven by standardization processes, platform emergence, regulatory interventions, or exogenous shocks such as pandemic-induced digital acceleration—remain unexamined. Future causal inference research could employ quasi-experimental designs leveraging policy interventions or technological breakthroughs as natural experiments, or apply instrumental variable approaches addressing endogeneity concerns inherent in observational patent data.

The limitations extend beyond methodological considerations to encompass substantive domains inadequately addressed in the current analysis. The study provides minimal discussion of ethical dimensions, sustainability implications, or societal consequences of IoT technology proliferation despite growing scholarly attention to algorithmic accountability, data sovereignty, environmental impacts of electronic waste, and digital divide exacerbation. Future research should integrate normative frameworks examining whether identified innovation trajectories align with sustainable development objectives, incorporate equity considerations in technology access and benefit distribution, and assess whether collaborative network structures perpetuate or mitigate existing power asymmetries in global innovation systems. The intersection of IoT technologies with emerging regulatory frameworks—including the European Union's Artificial Intelligence Act, data localization requirements, and carbon accounting mandates—represents critical domains requiring interdisciplinary inquiry bridging technology forecasting with policy analysis and sociotechnical systems scholarship.

## 6 Conclusion

This study advances understanding of Internet of Things innovation dynamics through integrated analysis of 154,227 patents, revealing nine distinct technology clusters exhibiting heterogeneous maturity profiles and demonstrating the multifaceted nature of IoT development. The application of logistic growth modeling quantifies technology life cycle positions, identifying imminent saturation in foundational domains including Mechanical Assembly and Hardware Architecture (2023) and Network Communication and Data Transmission (2025), while Data Management and Intelligent Platform Systems maintains growth trajectory toward 2027 saturation with the largest carrying capacity among all clusters. The remarkable temporal convergence of inflection points during 2020–2022 across technologically diverse domains provides empirical evidence of ecosystem-wide synchronization effects transcending individual cluster characteristics, likely catalyzed by pandemic-driven digital transformation imperatives and 5G infrastructure deployment.

Social network analysis of collaboration patterns reveals distinct strategic orientations among identified communities, ranging from extreme specialization (Global Telecommunications Technology Leaders: 87.7% network communication focus) to diversified portfolios addressing comprehensive infrastructure requirements (China State Grid IoT Consortium: balanced investment across monitoring, power management, cybersecurity, and platform domains). The cross-analysis connecting community membership with technological specialization patterns uncovers complementary innovation strategies where infrastructure operators pursue portfolio breadth, telecommunications specialists maintain focused expertise in enabling technologies, and academic researchers emphasize development-oriented agendas aligned with regional priorities. These findings illuminate the division of innovative labor within IoT ecosystems, demonstrating how heterogeneous organizational capabilities collectively enable comprehensive technological advancement.

The methodological contribution integrates state-of-the-art natural language processing through BERT-based patent embeddings with unsupervised clustering, S-curve life cycle modeling, and Louvain community detection within a unified analytical framework, providing a replicable template for technology forecasting in complex, rapidly evolving domains. The identification of high modularity network structures (Q = 0.963) combined with small-world topological properties (high clustering coefficient: 0.855; moderate average path length: 5.754) reveals efficient knowledge diffusion mechanisms within localized communities alongside global connectivity enabling cross-domain integration. For practitioners, the dual-perspective analysis linking technological maturity assessments with collaborative network positioning enables data-driven strategic planning, resource allocation optimization, and partnership formation decisions grounded in empirical understanding of innovation ecosystem dynamics rather than speculative projections or isolated technical assessments.

**Author contributions**

Mehrdad Maghsoudi: Conceptualization, Methodology, Data curation, Formal analysis, Patent data analysis, Visualization, Writing – original draft.Reza Nourbakhsh: Literature review, Data curation, Formal analysis, Patent data analysis, Visualization, Writing – original draft, Writing – review & editing.Mehrdad Agha Mohammadali Kermani: Conceptualization, Methodology,, Validation, Writing – review & editing.Rahim Khanizad: Writing – review & editing, Critical revision of the manuscript.

**Funding**

No funds, grants, or other support was received for conducting this study.

**Data availability**

The datasets used and analysed during the current study are available from the corresponding author on reasonable request.

## Declarations

**Ethics approval and consent to participate**

Not applicable.

**Consent for publication**

Not applicable.

**Competing interests**

The authors declare no competing interests.

## Publisher's note